\documentclass[onecolumn, authoryear]{elsarticle}
\pdfoutput=1

\usepackage[colorlinks=true,linkcolor=black, citecolor=blue, urlcolor=blue]{hyperref}

\usepackage{graphicx}
\usepackage{enumitem}
\usepackage{float}
\usepackage{cancel}
\usepackage{color}
\usepackage[a4paper, total={6in, 8in}, margin = 2.5cm]{geometry}
\usepackage{caption}
\usepackage{subcaption}
\usepackage{setspace}

\usepackage{psfrag,epsfig,graphics}
\usepackage{amsmath,amsthm,amssymb,multirow,mathrsfs}
\usepackage{mathbbol}
\usepackage{amssymb}             

\DeclareSymbolFontAlphabet{\amsmathbb}{AMSb}%
\usepackage{mathabx} 

\usepackage{booktabs}
\usepackage{graphicx,amsmath,graphicx}
\usepackage{caption}
\usepackage[position=b]{subcaption}

\usepackage{multirow}
\usepackage[lined,linesnumbered,ruled]{algorithm2e}

\usepackage{color}  
\def\cred{\textcolor{black}}
\def\credreview{\textcolor{black}}

\newcommand{\cp}[1]{\ifmmode {\mathcal{#1}}\else ${\mathcal{#1}}$\fi}

\newcommand{\bZ}{\boldsymbol{Z}}

\newcommand{\bz}{\boldsymbol{z}}

\usepackage[euler]{textgreek}
\usepackage[colorinlistoftodos]{todonotes}

\usepackage{color}  

\definecolor{darkgreen}{rgb}{0., 0.4, 0.}
\definecolor{amber}{rgb}{1.0, 0.49, 0.0}
\definecolor{orange}{rgb}{1.0, 0.4, 0.0}
\definecolor{darkorange}{rgb}{0.7, 0.32, 0.0}

\usepackage{float}
\usepackage{url}
\usepackage{makecell} 
\usepackage{siunitx} 
\usepackage{lineno}
\usepackage{xcolor}
\usepackage{colortbl}

\usepackage[title]{appendix}
\newcommand{\reffigure}[1]{\hyperref[#1]{Fig.~\ref{#1}}}
\newcommand{\refmultiplefigures}[2]{Figs.~\ref{#1} and~\ref{#2}}
\newcommand{\refmultiplefiguresthree}[3]{Figs.~\ref{#1}, ~\ref{#2} and~\ref{#3}}

\newcommand{\reftable}[1]{\hyperref[#1]{Table~\ref{#1}}}
\newcommand{\refsection}[1]{\hyperref[#1]{Section~\ref{#1}}}
\newcommand{\refequation}[1]{Eq.~\hyperref[#1]{(\ref{#1})}}
\newcommand{\refmultipleequations}[2]{Eqs.~\hyperref[#1]{(\ref{#1})} and~\hyperref[#2]{(\ref{#2})}}

\definecolor{atomictangerine}{rgb}{1.0, 0.6, 0.4} 

\newcommand{\refappendix}[1]{\hyperref[#1]{Appendix~\ref{#1}}}

\journal{ISPRS Journal of Photogrammetry and Remote Sensing}

\begin{document}
\begin{frontmatter}

%
\title{Recursive classification of satellite imaging time-series: \\ An application to land cover mapping}
%

\author[inst1]{Helena Calatrava\fnref{fn1}\corref{cor1}}
\author[inst1]{Bhavya Duvvuri\fnref{fn1}}
\author[inst1]{Haoqing Li}
\author[inst2]{Ricardo Borsoi}
\author[inst1]{Edward Beighley}
\author[inst1]{Deniz Erdo{\u{g}}mu{\c{s}}}
\author[inst1]{Pau Closas}
\author[inst1]{Tales Imbiriba}

\affiliation[inst1]{organization={Northeastern University},
            city={Boston},
            postcode={02215}, 
            state={MA},
            country={USA} }

\affiliation[inst2]{organization={CRAN, University of Lorraine, CNRS},
            city={Vandoeuvre-les-Nancy},
            postcode={F-54000}, 
            country={France}}
\cortext[cor1]{Corresponding author. \\ Email address: calatrava.h@northeastern.edu}
\fntext[fn1]{Indicates shared first authorship.}

\date{July 2023}

%
%
\begin{abstract}
\credreview{
Despite the extensive body of literature focused on remote sensing applications for land cover mapping and the availability of high-resolution satellite imagery, methods for continuously updating classification maps in real-time remain limited, especially when training data is scarce. 
This paper introduces the recursive Bayesian classifier (RBC), which converts any instantaneous classifier into a robust online method through a probabilistic framework that is resilient to non-informative image variations.
Three experiments are conducted using Sentinel-2 data: water mapping of the Oroville Dam in California and the Charles River basin in Massachusetts, and deforestation detection in the Amazon. RBC is applied to a Gaussian mixture model (GMM), logistic regression (LR), and our proposed spectral index classifier (SIC).
Results show that RBC significantly enhances classifier robustness in multitemporal settings under challenging conditions, such as cloud cover and cyanobacterial blooms.
Specifically, balanced classification accuracy improves by up to 26.95\% for SIC, 12.4\% for GMM, and 13.81\% for LR in water mapping, and by 15.25\%, 14.17\%, and 14.7\% in deforestation detection. 
Moreover, without additional training data, RBC improves the performance of the state-of-the-art DeepWaterMap and WatNet algorithms by up to 9.62\% and 11.03\%.
These benefits are provided by RBC while requiring minimal supervision and maintaining a low computational cost that remains constant for each time step regardless of the time-series length.
}
\end{abstract}



\end{frontmatter}



%
%
\section{Introduction}
\label{section:introduction}

\subsection{Background}\label{sec:intro:background}
For the purpose of allaying increasing concerns on global environmental changes and sustainability, and thanks to the vast amount of high resolution remotely sensed data available today, \credreview{there exists a considerable body of work focused on remote sensing applications involving land cover mapping and change detection~\citep{mashala2023systematic,wang2023review}.}
%
Some examples are studies on land conservation, sustainable development, and the management of resources such as water. Changes in water dynamics can be studied by surface water mapping to \credreview{monitor floods~\citep{yaseen2024flood}, describe water quality~\citep{wasehun2024uav}, and for coastline extraction~\citep{sun2023coastline}.}
%
Land cover mapping also plays a crucial role in identifying the distribution of \credreview{different crop types~\citep{ zhang2024crop} and understanding the dynamic evolution of land use in urban environments~\citep{yu2023urban}.}
%
%
%
\credreview{Furthermore, many studies focus on deforestation detection in areas such as the Amazon rainforest, a region with unparalleled biodiversity and a crucial role in global climate regulation, yet facing alarming deforestation rates. Numerous works address the environmental impact that this ongoing crisis has on climate change and public health~\citep{lapola2023drivers,ellwanger2020beyond}.} 
\credreview{Of particular concern is the 129\% increase in deforestation within Indigenous territories since 2013~\citep{silva2023brazilian}.}
%
%
%
Deforestation detection conducted by human experts through visual inspection is time-consuming and costly due to the vast geographic areas involved, \credreview{making automated detection methods necessary~\citep{isprs-annals-X-1-W1-2023-835-2023,MARTINEZ2024110}.}

\credreview{Several sources of remotely sensed data are currently available, presenting different characteristics when it comes to spatial, spectral, radiometric and temporal resolution~\citep{chuvieco2020fundamentals}.}
Spatial resolution varies from centimeters, as with high-resolution sensors on GeoEye and QuickBird-2 satellites, to a few meters, as with Landsat 9 and Sentinel-2 A/B sensors. These satellites can acquire weekly images of the same scene. Conversely, satellites with MODIS and VIIRS sensors offer daily image acquisitions but at a lower spatial resolution of hundreds of meters. \credreview{Modern commercial satellite systems push the boundaries of Earth observation~\citep{miura2023utility}.} An example is the Pléiades Neo by Airbus, with a twice-daily revisit capacity and a ground sampling distance of up to 30 cm~\citep{2022ISPAr43B1}.
Notably, hundreds of CubeSats with subweekly or even daily temporal resolution and medium to high spatial resolution~\citep{2022_survey}, \credreview{such as the ones providing Planetscope data~\citep{2021SciRS}.} Also, very high resolution optical imagery is available with the constellations of SkySat, BlackSky and Nu-Sat micro-satellites. The SkySat satellites can provide ground sampling distances of up to 50~cm and a sub-daily revisit time (6-7 times per day) when considering the whole constellation~\citep{isprs_skysat}.
Additionally, multimodal image fusion techniques are used to generate high spatio-temporal image sequences, contributing to generating a wealth of remotely sensed data~\citep{li2022onlineImFusionKalman, karmakar2023crop}.

\credreview{Spectral indices are one of the main land cover mapping tools given their simplicity and required low computational cost~\citep{tran2022review}.} They compute scalar-valued features as a function of specific spectral bands, whose value can be used to distinguish between different land cover classes contained in a pixel. 
Although they show a limited performance when compared to other techniques such as deep learning methods, they are widely used in remote sensing applications given their unsupervised nature~\citep{KHALID2021619}. They can be considered to be unsupervised because their output depends on the ratio between a combination of spectral bands, which does not require any training.
The decision threshold, however, must be selected and this can be challenging when no reference data is available.
Another advantage is that spectral index values are easy to interpret because they minimize the effect of illumination in satellite imagery while enhancing different spectral features present in the scene under study. \credreview{For instance, the normalized difference vegetation index (NDVI) enhances the presence of trees, bushes, and others~\citep{huang2021commentary, rs14163967}.} \cred{This is due to the reflectance given by the spectral response of vegetation decreasing in the red and increasing in the infrared wavelengths}. On the other hand, water indices are used for water extraction at pixel level, \credreview{given the difference in spectral reflectance of land and water in the near and middle infrared wavelengths \citep{liu2023remote}.} The most widely used water indices are the normalized difference water index (NDWI), the modified NDWI (MNDWI) and the automated water extraction index (AWEI)~\citep{water_indices}. \credreview{It has been shown that challenging weather conditions may disrupt the extraction of water bodies with these indices~\citep{yang2022review}.} This can be solved by using modified methods like the one proposed by~\citet{modified_method}, which effectively mitigates mountain shadows and allows the extraction of particularly challenging small water bodies. \citet{KHALID2021619} also suggest that the land surface temperature based water extraction index (LBWEI) provides high accuracy under a wide variety of weather conditions. 
%

%

%
Aside from spectral index methods, there is a wide choice of land cover classification approaches based on machine learning available in the literature, \credreview{whose main advantage is an increased flexibility~\citep{wang2022machine}.}
The taxonomy of image classification techniques in remote sensing proposed by~\citet{landclass_techniques_2012_table} groups them into supervised/unsupervised, parametric/non-parametric and hard/soft classifiers, among others. 
Some of the explored machine learning methods used in remote sensing applications include maximum likelihood classifiers \citep{zeb2019forest}, \credreview{support vector machines \citep{kok2021support}, logistic regression (LR) \citep{li2024incorporating}, random forests \citep{Pelletier2017,tariq2023modelling}, naive Bayes~\citep{bai2023naive} and clustering methods like the widely used K-means algorithm~\citep{ghezelbash2023genetic}.}
\credreview{\citet{liko2024deep} showed an improvement in maximum likelihood classification maps of Himalayan regions through post-classification correction measures, including ancillary data, digital elevation models, and spectral vegetation indices.} \credreview{Various studies assess the need for a comprehensive comparison between widely used machine learning algorithms for land cover classification~\citep{adugna2022comparison,zhang2023forest}. \citet{2022_ml} propose random forest as the best classification model in mining districts when compared to maximum likelihood, support vector machines, and classification and regression trees.}
Furthermore, the authors in~\citet{QIU2019151} suggest that deep learning methods like artificial neural networks provide high accuracy results in land cover classification, even when compared to other machine learning classifiers such as support vector machines.

Despite their widespread use in land cover mapping, the previously mentioned techniques suffer from several limitations. First, they are highly sensitive to illumination and atmospheric interferences (e.g., different aerosol concentrations or viewing angles), which can significantly impact the spectra of pixels from a given material class~\citep{borsoi2020variabilityReview}. The lack of robustness to such non-informative spectral variations is a significant limitation of, for instance, spectral indices such as the MNDWI~\citep{yang2018urbanWaterIndicesNoise}.
Moreover, due to the high sensor-to-target distances involved in remote sensing applications, many image pixels do not belong to a single class, \credreview{but are instead composed of a mixture of different material classes~\citep{cavalli2023spatial}. Although this can be addressed by spectral mixture analysis techniques~\citep{doi:10.34133/remotesensing.0117} or by assigning a pixel to more than one class with sub-pixel mapping~\citep{WANG2020111817,HU2021112365}, it poses a significant challenge to traditional classification algorithms. As a consequence, an apparent need for robustness to outliers and spurious artifacts in remotely sensed data arises. }

\subsection{Multitemporal time-series classification}

\credreview{The analysis of multitemporal or time-series data is of increasing interest for remote sensing applications~\citep{10529247}. Exploiting multitemporal data makes it possible to improve the performance of tasks such as classification \citep{lulc_time_Series,rs15133212} and spectral mixture analysis \citep{borsoi2021fastUnmixingChangeDetection} given the temporal correlation, while at the same time supplying the end-user with a more complete product that shows the spatial as well as the temporal distribution of land classes or their proportions.}
The simplest approach to perform multitemporal land cover mapping is to apply an instantaneous classifier to each image in the sequence, being spectral indices such as the NDVI a popular choice~\citep{sun2018classificationSaltMarshTimeSeriesNDVI}. However, this does not exploit the temporal information available in the data.
Significant effort has been dedicated to developing techniques specifically suited to process multitemporal image sequences. For instance, \citet{Kenduiywo2017} \credreview{and~\citet{9259004} proposed classification methods based on conditional random field models, which represent the interactions between class labels in both time and space.} Transfer and active learning have been combined to adapt a pre-trained classifier \credreview{to new images acquired at other time instants~\citep{8978543}.} Time-series classification accounting for missing pixels using Gaussian process regression was proposed by~\citet{Constantin2022}, while other works considered 1D temporal convolutional neural networks (CNNs)~\citep{Pelletier2019}, and 3D spatio-temporal CNNs~\citep{ji2018_3D_CNNs_time_crop_classification}.

%
These techniques, known as \emph{batch} or \emph{offline} time-series classification methods, require a complete image time-series to generate classification maps. However, with satellites like Sentinel-2 and Landsat~9 continuously acquiring images, reprocessing the entire series each time a new image is acquired becomes computationally expensive.
%
\textit{Recursive} methods, also known as \textit{online} methods, iteratively update multitemporal classification maps with new images by leveraging previously computed results, making them ideal for studies involving ongoing data collection.
%
Bayesian recursion is widely used in target tracking applications, where an efficient filter~\citep{imbiriba2020enhancing} must be designed to recursively obtain target state estimates from a state-space model~\citep{Ji2022RecursiveBI_tracking}. Filtering methods can also be applied to online parameter learning~\citep{wu2019wifi, borsoi2020kalman, demirkaya2021cubature, imbiriba2022hybrid}, which is of special interest in machine learning tasks involving the processing of time-series data~\citep{campbell2021online}, such as video prediction~\citep{pmlr-v119-franceschi20a} and speech enhancement~\citep{2020_speech_enhancement}. Bayesian recursion has also been applied in pattern recognition for bioengineering applications, such as tracking retinal vasculature by estimating vessel geometry parameters~\citep{2019_retinalrecursive}. 

%

The earliest recursive remote sensing classification techniques were based on Bayesian filtering ideas, by recursively updating the probabilities of each class given the measurements after each datum is acquired~\citep{Swain1978, Strahler1980}. These techniques are based on a statistical model that represents the pixel spectra given its class, \credreview{called a \emph{generative model}, which is non-trivial to obtain. More recent Bayesian approaches have proposed classification strategies that are recursive both in time and across multiple spatial scales (multiresolution)~\citep{2021_hierarchical},} using computationally expensive algorithms such as expectation maximization to learn the model parameters and a generative model for the pixels. Other recent works have leveraged deep learning strategies, in particular different instantiations of recurrent neural networks, such as \credreview{long short-term memory (LSTM) networks~\citep{ZHOU2023110394,rs16020324}.} These have been applied to predict flood susceptibility and for crop identification, among others~\citep{Rubwurm2017, Fang2021}.
These methods enable learning intrinsic spatial and temporal dependencies of remotely sensed data, along with spectral patterns in specific classes over time, all with minimal supervision.
%
The main disadvantage of other recursive algorithms is that they require large amounts of training data and long training times compared to the framework proposed in this paper. This is the case for the algorithm in~\citet{sharma2018landCoverClassificationPatchRNNs}, which uses \credreview{patch-based recurrent neural networks, and the one in~\citet{10557465}, which combines fully convolutional networks with hierarchical probabilistic graphical models and decision tree ensembles.}
\citet{2023JPRS} show that deep learning methods provide substantial classification improvements over the commonly implemented random forest. Nevertheless, demonstrative numbers on simulation times suggest that these methods require up to four times larger running times.

\credreview{The lack of open-source labeled Sentinel-2 data for time-series analysis (i.e., containing the dynamical evolution of the true class labels) poses an important challenge to the evaluation of the methods. 
Although there are available multitemporal satellite imagery datasets, most of them contain pixel-wise annotations that are unique for the whole time-series. That is, no changes over time exist or are properly mapped to labels, and therefore, it is not possible to evaluate the algorithm at different time instants. One example of this is the benchmark dataset for multi-temporal and multi-modal land use land cover mapping MultiSenGE~\citep{multisenge_dataset}. To address this, we generated our own ground truth data, as detailed in the next subsection. For the Amazon deforestation experiment, ground truth labels were available in the MultiEarth challenge dataset~\citep{cha2023multiearth}.}
%

%
%
%
\begin{figure*}[th]
    \hspace{-0.55cm}
    \includegraphics[width=1.08\linewidth]{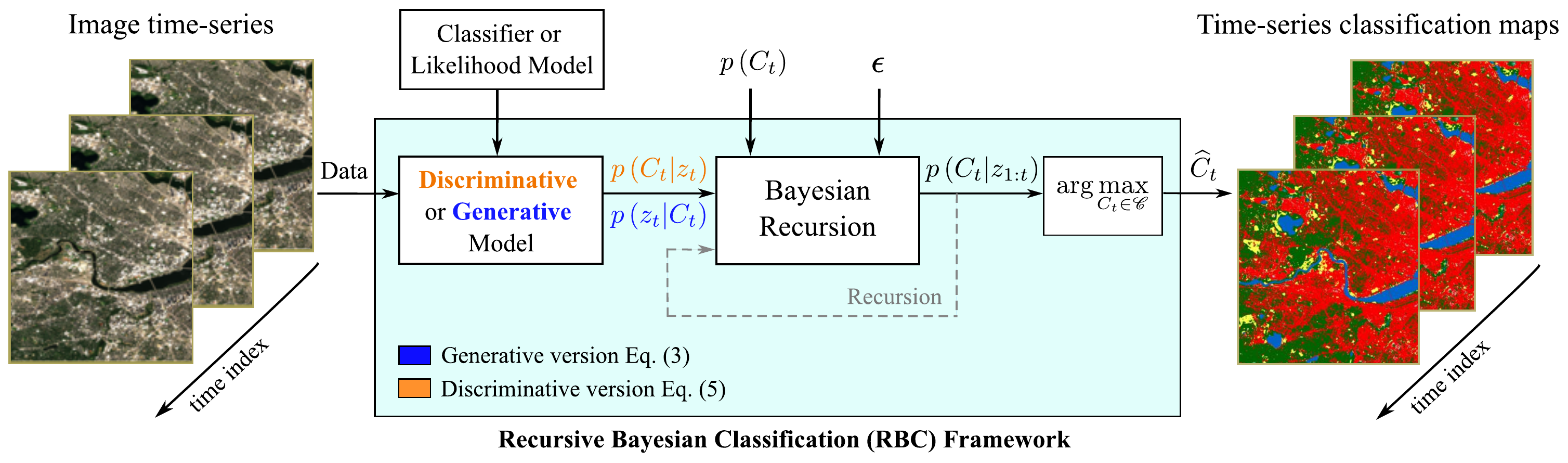}
    \caption{Overview of the proposed recursive Bayesian classification framework. Our method allows to convert an instantaneous generative model (i.e., modeling the observation of a pixel given its class label with the likelihood function $p\left(\bz_t|C_t\right)$) or a discriminative model (i.e., modeling the observation of a class label given a corresponding pixel with $p\left(C_t|\bz_t\right)$) into a recursive Bayesian classifier that exploits the temporal relationship of time-series data. Knowledge about the class prior probabilities $p\left(C_t\right)$ and the \credreview{transition probability hyperparameter $\epsilon$} is assumed. The hat operator in $\widehat{C}_t$ denotes the decision from the classifier. $\mathscr{C}$ is an experiment-dependent set containing $K$ different labels.}
    \label{fig:graphical_abstract}
\end{figure*}

\subsection{Contributions}
\credreview{
This paper aims to develop a recursive classification framework that improves the decision-making process in multitemporal and multispectral land cover classification algorithms by leveraging previous classification results. An overview of the proposed framework may be found in~\reffigure{fig:graphical_abstract}.
The key contributions of this paper are as follows:
\begin{itemize}
    \item We propose the recursive Bayesian classification (RBC) framework, which converts any instantaneous classifier into a robust online method using a probabilistic framework that enables the integration of temporal information. RBC addresses the trade-off between adaptability to natural changes in the scene and robustness to outliers caused by illumination variations and spurious artifacts in remotely sensed data. This is achieved without the need for additional training data, as opposed to more complex deep learning models.
    \item In~\refsection{section:recursive_spectral_index_classification}, we introduce the spectral index classification (SIC) algorithm, which uses standard broadband spectral indices to generate predictive probabilities. This approach allows for the incorporation of class prediction uncertainty into the RBC framework.
    \item 
    To overcome the lack of available ground truth data for time-series analysis, we created our own ground truth data with manually generated labels. This enabled a rigorous quantitative assessment of the proposed framework (see \refsection{sec:evaluation}). The ground truth dataset, along with the pre-processed satellite imagery, has been made publicly available at~\citep{zenodo_link}.
\end{itemize}
%
}

The performance of the proposed approach is demonstrated using Sentinel-2 images in three different experiments: water mapping of a reservoir and downstream river in Oroville, California, USA; water mapping of the Charles river basin in Boston, Massachusetts, USA; and deforestation detection in the Amazon rainforest. RBC may be applied atop a generative model (see~\refsection{sec:rbcg}) or atop discriminative models (see~\refsection{sec:rbcd}). We evaluate the proposed RBC framework against three instantaneous classifiers: a Gaussian mixture model (GMM), an LR classifier, and the SIC algorithm, along with their recursive counterparts. For the water mapping experiments, we include two state-of-the-art deep learning classifiers: the DeepWaterMap (DWM)~\citep{8013683} and WatNet (WN)~\citep{luo2021_watnet} algorithms. 

The remainder of this paper is structured as follows. Section \ref{section:data_and_aos} describes the satellite data and the are under study. Section \ref{section:methods} introduces the proposed RBC framework and the SIC algorithm, followed by details on the experimental setup. Section \ref{section:results} presents the results, while their implications and impact are discussed in Section \ref{section:discussion}. Finally, Section \ref{section:conclusion} provides the concluding remarks. \cred{The successful extension of the RBC framework to a three-class classification task for the Charles River area is detailed in Appendix~\ref{appendix:charles_river}.}

\section{Area of study and satellite data}
\label{section:data_and_aos}
%
%
%

Our research focuses on two areas of study located in the US and one area located in the Amazon rainforest, \cred{as we believe it is essential to test our methodology across varied geographical locations to achieve a more comprehensive performance assessment. In this section, we describe the three selected areas of study and the challenges they pose to land cover mapping algorithms.}
\begin{figure}[H]
    \centering
    \includegraphics[width=15cm]{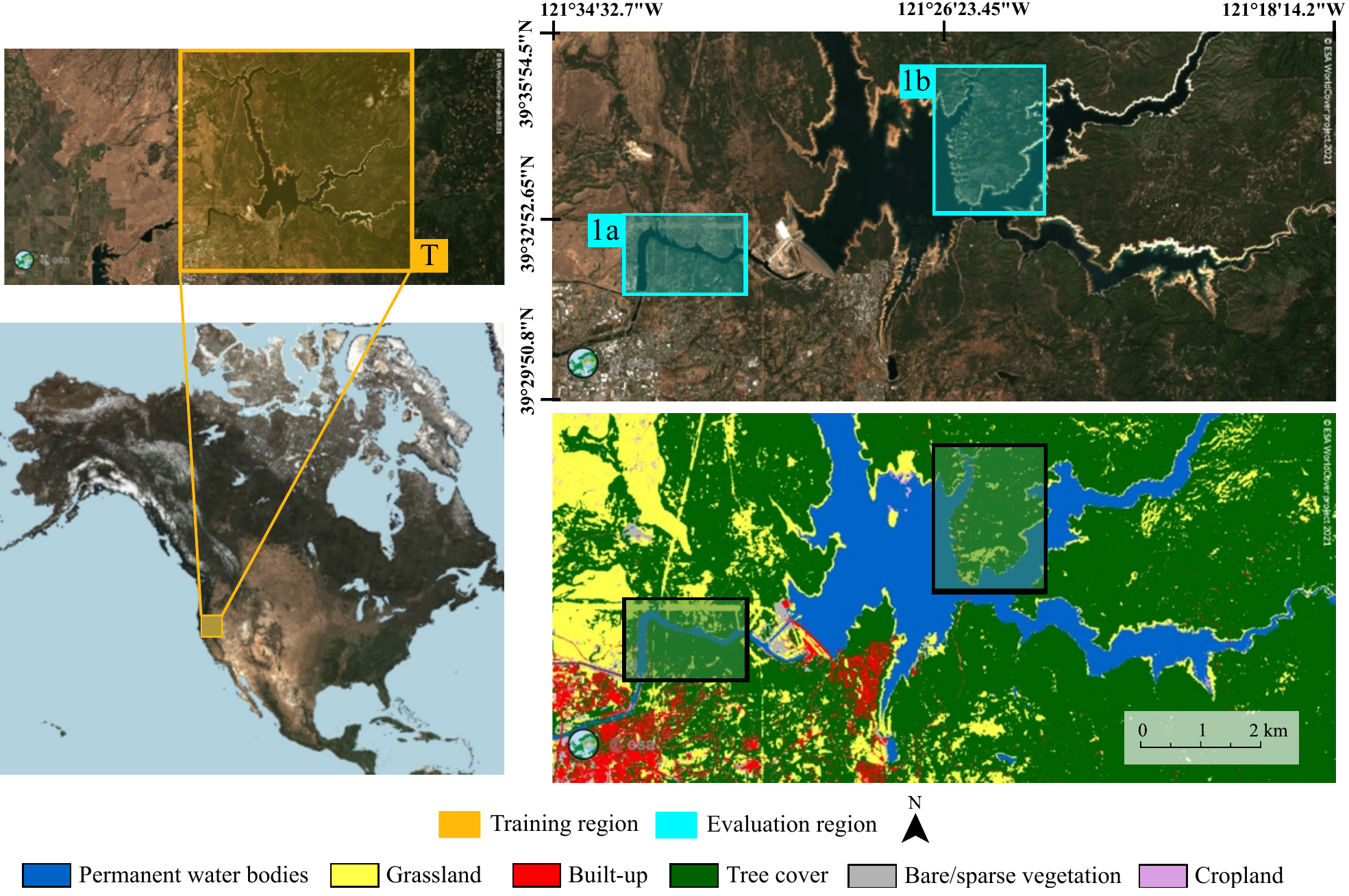}
    \caption{Sentinel-2 RGB composite image \credreview{with LAT/LON information} and classification map from the ESA WorldCover Map tool~\citep{esa_world_cover_map} of test sites 1a and 1b (Oroville Dam, California, USA).}
    \label{fig:data_aos:areas_of_study:oroville_dam}
\end{figure}

We consider images at the blue, green, red, near-infrared (NIR), narrow NIR and shortwave infrared (SWIR) bands, with resolution and central wavelength listed in~\reftable{tab:data_aos:sentinel-2_bands}. Band 4 (red) is useful for identifying soil, water and many urban features, band 3 (green) gives excellent contrast between clear and turbid waters, and band 2 (blue) is useful for identifying vegetation and also human-made features~\citep{Huk_Maleszka_Szczerbicki_2020}. The SWIR bands are useful for measuring vegetation, water and soil moisture.

\begin{table}[H]
    \centering
    \caption{Resolution and central wavelength of the processed Sentinel-2 spectral bands.}
    \begin{tabular}{c c c c}
        \hline
        Band & Description & Resolution (m) & Central wavelength (nm) \\ \hline 
        2 & Blue & 10 & 490\\
        3 & Green & 10 & 	560\\
        4 & Red & 10 & 665\\
        8 & NIR & 10 & 842\\
        8A & Narrow NIR Edge & 20 & 	865\\
        11 & SWIR 1 & 20 & 1610\\
        \hline
    \end{tabular}
    \label{tab:data_aos:sentinel-2_bands}
\end{table}

\subsection{Test site 1: Oroville Dam}
The first test site is located in Oroville Dam, an embankment dam on the east side of the city of Oroville, in the state of California (see~\reffigure{fig:data_aos:areas_of_study:oroville_dam}). Being 235 meters high, it is the tallest dam in the US. The area of study has geographic center coordinates of LAT/LON: 39.61, -121.43.  Test sites 1a and 1b belong to the dam downstream and upstream, respectively. The water mapping of areas with geographic features like this reservoir is imperative to study, as they are essential for flood control, management and sustainability of water resources.

Illumination variations in images from test site 1a can be observed in~\reffigure{fig:data_aos:challenges}. These may be caused by fluctuations in the solar incidence angle and also by differences in image acquisition times. Ripples and other artifacts in test site 1a are caused by the high flows in the river stream, making the classification of water pixels challenging and increasing the probability of false negatives.
As illustrated in~\reffigure{fig:data_aos:challenges} for test site 1b, and based on reservoir storage data obtained from the NWIS USGS website~\footnote{\url{https://waterdata.usgs.gov/nwis/}}, the water level in the reservoir changes abruptly with the season.
In October 2020, the recorded water storage was of 200,485.8 hc-m, decreasing to 160,783.2 hc-m in December 2020. Subsequent changes resulted in a recorded water storage of 183,223.8 hc-m and 97,542.6 hc-m in May and September of 2021, respectively. 
These phenomena demand the flexibility of the proposed recursive classification framework, which ensures its ability to adapt to changes in the scene. However, the very high flexibility of the method can put its robustness at risk.

Sentinel-2 Level-2A images are downloaded using the Google Earth Engine platform from the COPERNICUS/S2\_SR collection, with dates between 2020-09-01 and 2021-09-26. 
This platform atmospherically corrects the images using the standard SEN2COR software package and indicates a cloud cover percentage.
Only images with at most 10\% cloud cover are downloaded. 
The downloaded images have dimensions of $2229 \times 3341$ pixels. For evaluation purposes, images were cropped to sizes of $200 \times 500$ pixels for test site 1a and $150 \times 110$ pixels for test site 1b. 
To ensure consistency, images from bands 8A and 11, initially at a resolution of 20 meters, were resampled to 10 meters using nearest-neighbor interpolation. A total of 45 images remained available for further processing.

\begin{figure}[H]
    \centering
    \includegraphics[width=15cm]{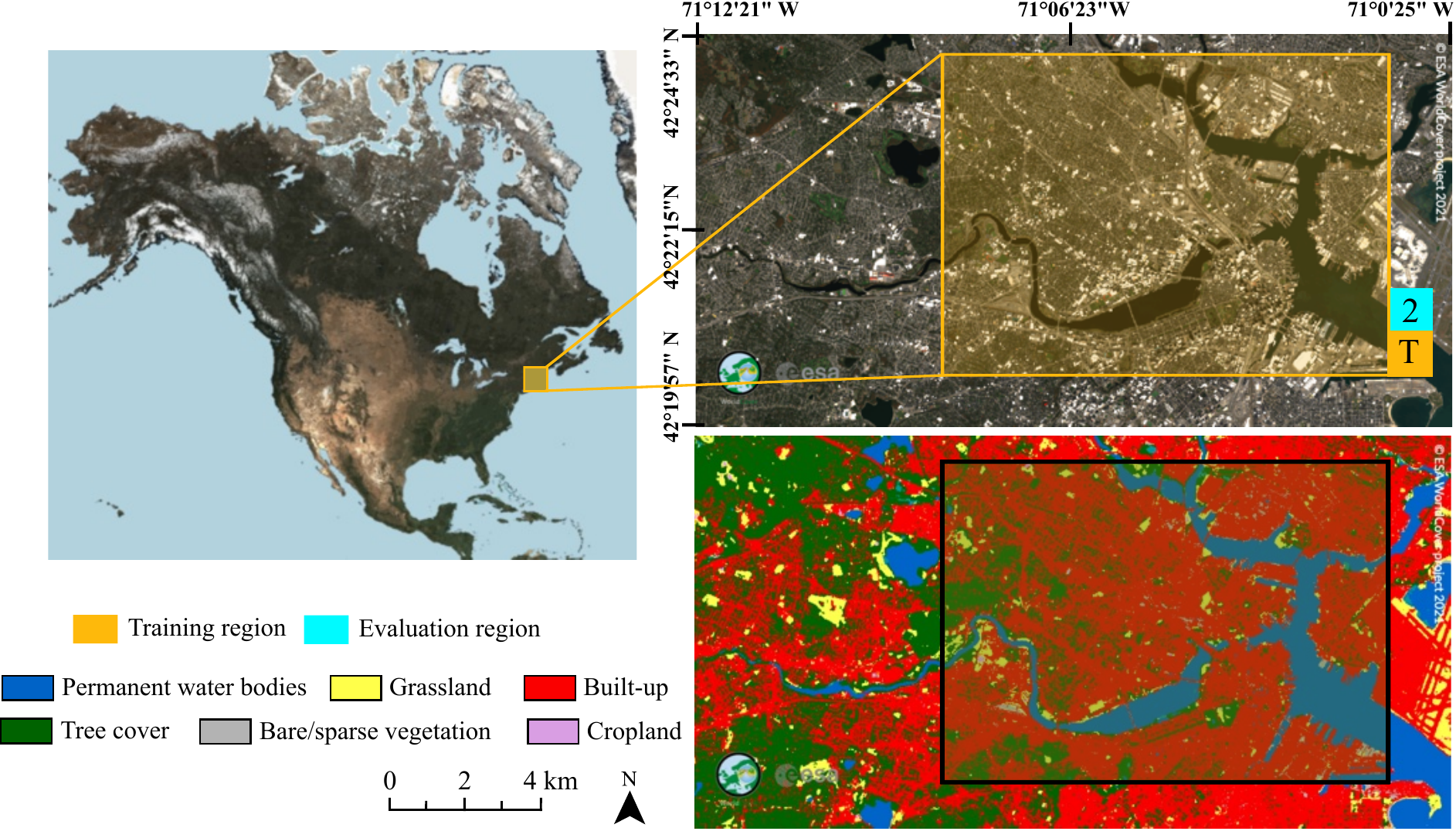}
    \caption{Sentinel-2 RGB composite images \credreview{with LAT/LON information} and classification map from the ESA WorldCover Map tool~\citep{esa_world_cover_map} of test site 2 (Charles river, Boston, USA).}
    \label{fig:data_aos:areas_of_study:charles_river}
\end{figure}
\subsection{Test site 2: Charles river basin}
\begin{figure}[h]
    \centering
    \includegraphics[scale=0.85]{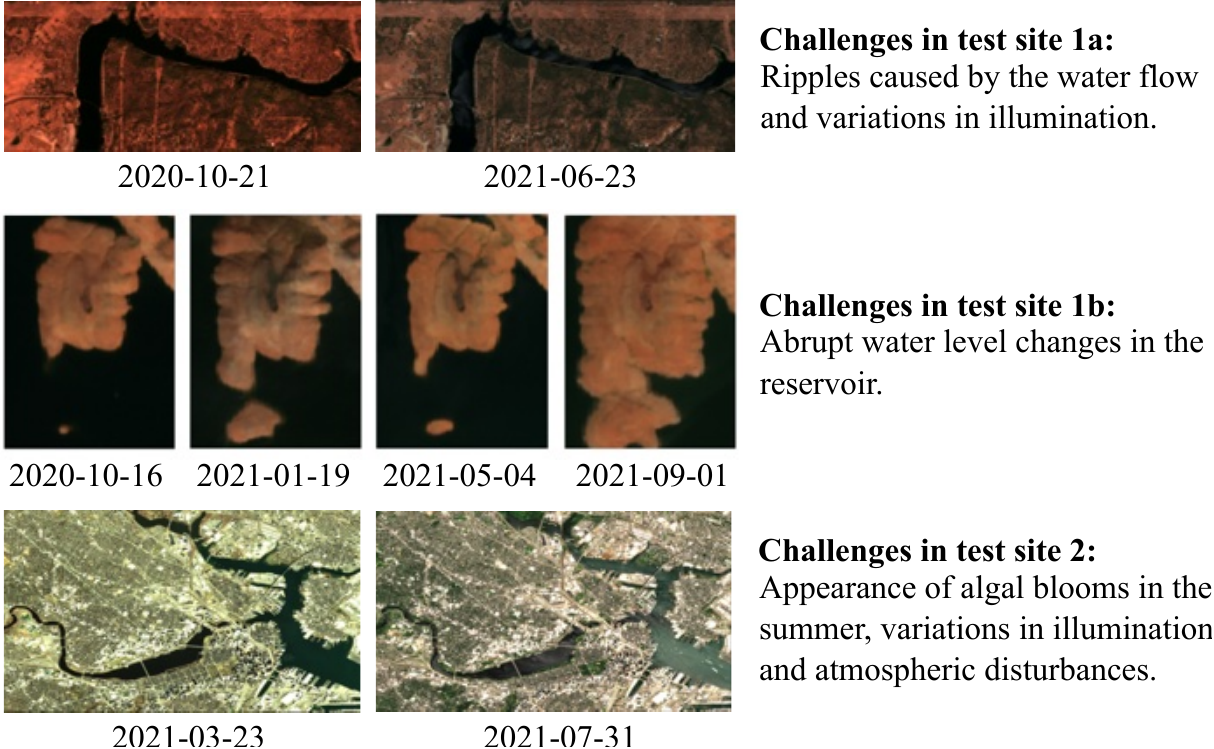}
    \caption{Sentinel-2 RGB composite images showing the challenges posed by test sites 1a, 1b, and 2, within the context of water mapping and land cover classification.}
    \label{fig:data_aos:challenges}
\end{figure}

The area of study for this experiment covers the Charles river, the Mystic river and the Boston harbor in Massachusetts, with geographic center coordinates of LAT/LON: 42.36, -71.12 (see~\reffigure{fig:data_aos:areas_of_study:charles_river}). This location includes a big permanent water body, urban vegetation, and a built-up area, which makes it a site of interest for land cover classification. Tracking land cover changes in urban environments is of great help for urban and agriculture planning, and also when trying to identify correlations between social activities and land changes.
Challenges in test site 2 include illumination variations (see~\reffigure{fig:data_aos:challenges}) and the presence of reflective surfaces from buildings and seasonal cyanobacterial blooms in the Charles river and in the Boston harbor waters. Algal blooms mostly occur during summer~\citep{rome2021sensor}, as in the image captured on 2021-07-31 shown in~\reffigure{fig:data_aos:challenges}.

Sentinel-2 Level-2A images are downloaded using the Google Earth Engine platform from the COPERNICUS/S2\_SR collection, with dates between 2020-09-04 and 2021-09-26.
Only images with at most 10\% cloud cover are downloaded.
The downloaded images have dimensions of $927 \times 2041$ pixels. For evaluation purposes, images were cropped to sizes of $700 \times 1241$ pixels.
Images from bands 8A and 11 are resampled to 10 meters by nearest-neighbor interpolation. After visual inspection, 15 images depicting snow-covered land or exhibiting significant disparities are excluded from the dataset, resulting in a total of 28 images for further processing.

\subsection{Test site 3: Amazon rainforest}
\begin{figure}[h]
    \centering
    \includegraphics[width=12.5cm]{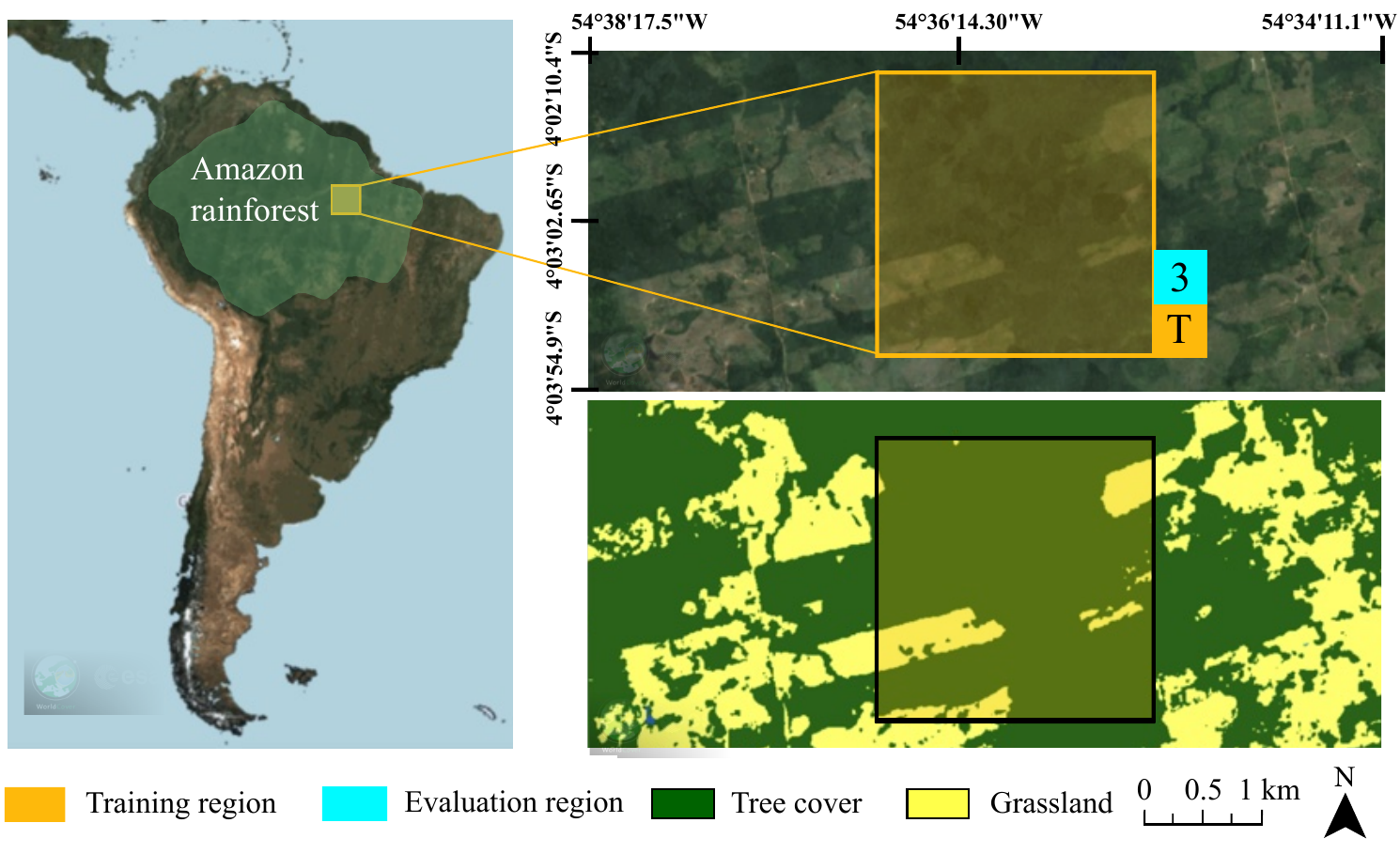}
    \caption{Sentinel-2 RGB composite images \credreview{with LAT/LON information} and classification maps from the ESA WorldCover Map tool~\citep{esa_world_cover_map} of test site 3, located in the endangered Amazon rainforest.}
    \label{fig:data_aos:areas_of_study:amazon}
\end{figure}

The third experiment focuses on the geographic area of the Amazon rainforest in Brazil, at the geographic center coordinates LAT/LON: -4.05, -54.6 (see~\reffigure{fig:data_aos:areas_of_study:amazon}).
Sentinel-2 images of the Amazon rainforest downloaded using the Google Earth Engine platform are obtained from the MultiEarth challenge dataset~\citep{cha2023multiearth}. For some dates, this dataset includes manually generated deforestation labels using mosaic satellite images from the Planet APIs\footnote{\url{https://api.planet.com/}}.
We initially considered a total of 225 images with dates between 2018-12-03 and 2021-12-27. These have the same dimensions as the MultiEarth dataset images after being tiled and segmented, i.e., $256\times 256$ pixels. The study area is selected to be relatively small in order to ease the analysis of results since the evolution in time of output classification maps can be more easily interpreted for smaller regions. Also, the overall computational cost decreases when lowering the number of pixels to be evaluated. 

The percentages of cloud and cloud shadow are calculated with the CI$_2$ and CSI indices proposed by~\citet{zhai_cloud_shadow_detector}. From the initially 225 selected images, the ones with a cloud/shadow cover above 20\% are filtered out. We also discard 12 images by visual inspection. This results in a total of 31 images for further processing.
Pixels in the images from the MultiEarth dataset are shifted accordingly such that surface reflectance values are between 0 and 1, showing a considerable bias under the presence of clouds or illumination factors. To solve this, a time-varying bias is fitted to each image, for which an area where the statistics are expected to be time-invariant (i.e., no clouds or disturbances are observed inside that area for all evaluation dates) is selected. Taking the first evaluation image as a reference, the mean $\bar{x}_\text{ref}$ of pixel values inside that area is calculated. For each subsequent image, a bias is computed as $b(t) = \bar{x}_\text{ref} - \bar{x}_\text{test}(t)$ and applied in the pre-processing stage so that the pixel mean in the selected region is the same for all time instants. With this procedure, pixel values corresponding to the surface reflectance are not altered by the presence of clouds or other atmospheric effects.

Even after the detection and filtering of images with a relatively high cloud percentage, scenes in the Amazon rainforest often show a large number of small clouds, which may disrupt instantaneous classification. On the one hand, this demands the robustness of the proposed RBC framework. On the other hand, the temporal variability of the spectra of both vegetated and, especially, deforested areas, requires flexibility in the algorithm. This variability can be observed in the Sentinel-2 RGB composite images from dates 2019-08-10 and 2020-06-10 in~\reffigure{fig:results:experiment_3_qualitative_analysis}. Overall, the proposed framework must address a trade-off between adaptability and robustness to surpass these challenges.

%

%
%
\section{Methodology}\label{section:methods}

The main contribution of this manuscript is the framework for recursive Bayesian classification using multispectral and multitemporal data, which is introduced in~\refsection{sec:algorithm}. We also propose, in~\refsection{section:recursive_spectral_index_classification}, a classifier that uses spectral indices to generate predictive class probabilities. The experimental setup is described in~\refsection{section:experimental_setup}.
%
\subsection{Algorithm: recursive Bayesian classification (RBC)}\label{sec:algorithm}
\begin{figure}[h]
    \centering
    \begin{subfigure}[b]{0.45\textwidth}  
        \centering
        \includegraphics[width=\textwidth]{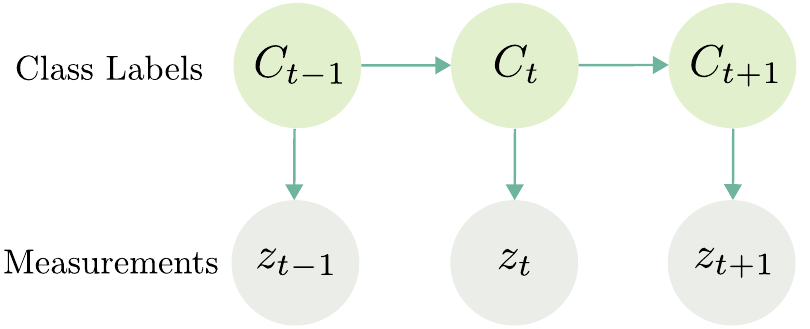}  
        \caption{}
        \label{fig:bayesian_recursion_a}
    \end{subfigure}
    \hfill
    \begin{subfigure}[b]{0.45\textwidth}  
        \centering
        \includegraphics[width=0.8\textwidth]{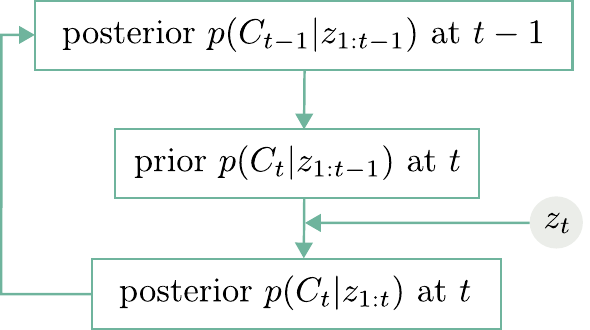}  
        \caption{}
        \label{fig:bayesian_recursion_b}
    \end{subfigure}
    \caption{\credreview{Diagrams illustrating the Bayesian recursion process. (a) Dependencies between class labels and pixel observations based on the first-order Markov assumption. (b) Bayesian recursion flow, updating the prior distribution to the posterior distribution using new pixel observations.}}
    \label{fig:bayesian_recursion}
\end{figure}

\credreview{The Bayesian philosophy involves updating beliefs based on evidence. It begins with a prior probability, i.e., the initial belief about a class label, which is updated with new data to generate a posterior probability, i.e., the refined belief after considering the new observation~\citep{sarkka2023bayesian}.}
%
%
%
%
Let us denote by $\bZ_{t}\in\amsmathbb{R}^{B\times N}$ an image with $B$ bands and $N$ pixels observed at time instant $t\in\{1,\ldots,T\}$. The images at the different time instants are supposed to be coregistered, that is, they constitute observations of the same geographic scene. For each pixel $\bz_{t,n}\in\amsmathbb{R}^B$, being $n\in\{1,\ldots,N\}$, we associate a label $C_{t,n}\in\mathscr{C}$, where $\mathscr{C}$ is an experiment-dependent set containg the possible $K$ labels.
\credreview{Bayesian recursion allows the refinement of class probabilities at each time step as new data is observed.} For a set of images $\bZ_t$ over time, the most likely label $C_{t,n}$ for each pixel $\bz_{t,n}$ (i.e., the $n$-th column of $\bZ_t$) can be estimated based on all the previous imagery $\{\bZ_t,\bZ_{t-1},\ldots,\bZ_1\}$ by maximizing the posterior probability $p(C_{t,n}|\bZ_t,\bZ_{t-1},\ldots,\bZ_1)$ as
\begin{equation}
    \label{eq:methods:general_classifier}
    \widehat{C}_{t,n} = \arg\max_{C_{t,n}\in \mathscr{C}} \,\, p(C_{t,n}|\bZ_t,\bZ_{t-1},\ldots,\bZ_1) \,,
\end{equation}
\credreview{where $\widehat{C}_{t,n}$ denotes the decision from the classifier.}
The expression in~\refequation{eq:methods:general_classifier} is powerful, as it considers both temporal and spatial information. However, learning the posterior PMF in~\refequation{eq:methods:general_classifier} can be hard, especially with high dimensional images. A spatial independence assumption can be applied to reduce the computational cost when calculating the conditional PMF. We propose to treat the label of every pixel as independent of the data from other pixels, meaning that $C_{t,n}$ only depends on $\bz_{t,n},\bz_{t-1,n},\ldots,\bz_{1,n}$, or, equivalently, on $\bz_{1:t,n}\triangleq\{\bz_{t,n},\bz_{t-1,n},\ldots,\bz_{1,n}\}$. This is without loss of generality, as the proposed approach can be directly extended to consider spatial information (i.e., from multiple pixels). Thus, the posterior in~\refequation{eq:methods:general_classifier} becomes $p(C_{t,n}|\bz_{1:t,n})$, disregarding spatial information, and leading to
%
\begin{equation}
    \label{eq:methods:classifier_only_temporal_dependence}
    \widehat{C}_{t} = \arg\max_{C_{t}\in \mathscr{C}} \,\, p(C_{t}|\bz_{1:t}),
\end{equation}
where the pixel index $n$ is omitted for simplicity. The classifier proposed in~\refequation{eq:methods:classifier_only_temporal_dependence} still considers a temporal dependence on previous data, meaning that the labels and images at previous time instants influence the results of the current time $t$. 
\credreview{Specifically, we assume first-order Markovity, with the dependencies between class labels (states) and pixel observations (measurements) depicted in~\reffigure{fig:bayesian_recursion_a}. This Markov property implies that $C_t$, (and the whole future $C_{t+1}$, $C_{t+2}$,...) given $C_{t-1}$ is independent of anything that occurred prior to time step $t-1$.}

\credreview{The probability of a class label \( C_{t-1} \) given all the data up to time \( t-1 \) corresponds to the posterior distribution at time \( t-1 \), denoted as $p(C_{t-1}|z_{1:t-1})$. From this distribution, the transition probabilities allow us to build the prior or predictive distribution at time step $t$ as $p(C_t|z_{1:t-1})$. Finally, with Bayes' rule, we can compute the posterior distribution with the newly observed pixel $z_t$ as $p(C_t|z_{1:t})$. This process allows the probability of class labels to be refined at each time step and is illustrated in~\reffigure{fig:bayesian_recursion_b}.}
With RBC, recursion can be applied to generative models (see~\refsection{sec:rbcg}), and to discriminative models (see~\refsection{sec:rbcd}).

\subsubsection{Classification based on a generative model (RBGM)}\label{sec:rbcg}
The posterior PMF in~\refequation{eq:methods:classifier_only_temporal_dependence} can be computed recursively using Bayes theorem under conditional independence assumptions and assuming knowledge about the \credreview{class prior probabilities $p(C_0)$, which reflect the initial understanding about the scene}, the state transition $p(C_t|C_{t-1})$ and the likelihood distribution $p(\bz_t|C_t)$ given by a generative model. Thus, the posterior PMF can be computed as
\begin{linenomath}
\begin{align}
    \label{eq:methods:posterior_generative_model_RBGM}
    \begin{split}
    p(C_t|\bz_{1:t})
    &\mathop{=}^{(a)}\frac{p(\bz_t|C_t)p(C_t|\bz_{1:t-1})}{p(\bz_t|\bz_{1:t-1})} \\
    &\mathop{=}^{(b)} p(\bz_t|C_t) \frac{\sum_{C_{t-1}\in\mathscr{C}} p(C_t|C_{t-1}) p(C_{t-1}|\bz_{1:t-1})}{\sum_{C_t'\in\mathscr{C}} p(\bz_t|C_t')p(C_t'|\bz_{1:t-1})}\\
    &
    = p(\bz_t|C_t) \frac{\sum_{C_{t-1}\in\mathscr{C}} p(C_t|C_{t-1}) p(C_{t-1}|\bz_{1:t-1})}{\sum_{C_t'\in\mathscr{C}} p(\bz_t|C_t') \sum_{C_{t-1}'\in\mathscr{C}}p(C_t'|C_{t-1}')p(C_{t-1}'|\bz_{1:t-1})}\,,
    \end{split}
\end{align}
\end{linenomath}


%
\noindent where in equality $(a)$ we assumed the conditional independence of measurements, that is, given the class label $C_t$, $\bz_t$ is independent of the previous $\bz_{1:t-1}$; in equality $(b)$ we assumed a first-order Markov model, considering that given the previous class label $C_{t-1}$, the current class label is independent of past measurements.
We refer to the method in~\refequation{eq:methods:posterior_generative_model_RBGM} as recursive Bayesian classification based on a generative model (RBGM) due to its dependence on the likelihood function $p(\bz_{t}|C_t)$.
The term $p(C_{t-1}|\bz_{1:t-1})$ denotes the posterior PMF of the previous time step. 
When $t=1$, $p(C_{t-1}|\bz_{1:t-1})=p(C_0)$ becomes equivalent to the marginal class probabilities at $t=0$, which we assume to be uniform in the absence of prior information about the scene, i.e., $p(C_0)=\frac1{K}\; \forall~C_t$.
%
The transition PMF $p(C_t|C_{t-1})$ is described later in this section.

%

%
%
\subsubsection{Classification based on a discriminative model (RBDM)}\label{sec:rbcd}
The posterior probability can also be calculated as a function of the probability of the labels given the pixel values, which allows existing classification algorithms to be used in the RBC framework. We refer to this as recursive Bayesian classification based on a discriminative model (RBDM). 
Applying \credreview{the Bayes' rule to the likelihood $p(\bz_{t}|C_t)$ we obtain}
\begin{linenomath}
\begin{align}
    p(\bz_{t}|C_t) = \frac{p(C_t|\bz_{t})p(\bz_{t})}{p(C_t)},
\end{align}
\end{linenomath}
where $p(C_t|\bz_{t})$ is the prediction of the classifier to which the RBC framework is applied, i.e., the benchmark classifier. As the RBC framework is agnostic to the classifier that is used, the prediction can be the result of any type of classifier, including deep learning methods as well. The Bayes theorem can be used to extend RBGM to RBDM by applying the Bayes theorem to~\refequation{eq:methods:posterior_generative_model_RBGM} as
%
\begin{linenomath}
\begin{align}
    \label{eq:methods:posterior_discriminative_model_RBDM}
    \begin{split}
    p(C_t|\bz_{1:t})
    &= 
     \frac{p(C_t|\bz_{t})\cancel{p(\bz_{t})}}{p(C_t)} \frac{\sum_{C_{t-1}\in\mathscr{C}} p(C_t|C_{t-1}) p(C_{t-1}|\bz_{1:t-1})}{\sum_{C_t'\in\mathscr{C}}  \frac{p(C_t'|\bz_{t})\cancel{p(\bz_{t})}}{p(C_t')} \sum_{C_{t-1}'\in\mathscr{C}}p(C_t'|C_{t-1}')p(C_{t-1}'|\bz_{1:t-1})}\\
     &=
     \frac{p(C_t|\bz_{t})}{p(C_t)} \frac{\sum_{C_{t-1}\in\mathscr{C}} p(C_t|C_{t-1}) p(C_{t-1}|\bz_{1:t-1})}{\sum_{C_t'\in\mathscr{C}}  \frac{p(C_t'|\bz_{t})}{p(C_t')} \sum_{C_{t-1}'\in\mathscr{C}}p(C_t'|C_{t-1}')p(C_{t-1}'|\bz_{1:t-1})}\,,
    \end{split}
\end{align}
\end{linenomath}
where $p(C_t)$ denotes the marginal class probability. In the widely used naive Bayes classifier, the marginal class probabilities are also used~\citep{barberBRML2011}. In the absence of labeled training data and prior information about the scene, we set their value to $p(C_t)=\frac1{K}\; \forall~C_t$.

Note that the proposed recursive classification solution is in closed-form and consists of a summation of probability distributions over the different classes. \refmultipleequations{eq:methods:posterior_generative_model_RBGM}{eq:methods:posterior_discriminative_model_RBDM} are straightforward to compute given the likelihood of the pixels or their posterior probability, respectively, which correspond to the result given by the instantaneous classifier. Considering this, it can be stated that recursion does not add a significant computational overhead to the classification problem.

\subsubsection{Class transition probabilities}\label{sec:class_transition_probabilities}
In this work, we assume that the class transition probability \( p(C_t | C_{t-1}) \), \credreview{which represents the likelihood of transitioning from class \( C_{t-1} \) to class \( C_t \)}, is time-invariant. Although strong, this assumption copes with the lack of knowledge we assume regarding the studied scene. Moreover, we highlight that this is without loss of generality since prior knowledge about, e.g., seasonality, can be easily incorporated in a time-dependent transition PMF. 
\credreview{The stationary or time-invariant case requires selecting \( K^2 \) parameters, but this is simplified by assuming \( p_{ij} = p(C_t = j | C_{t-1} = i) = \epsilon \) for all \( i \neq j \) when $K=2$, reducing the problem to selecting a single parameter. This parameter, known as the transition probability hyperparameter \( \epsilon \), corresponds to the probability of a pixel transitioning from one label to another in the two-class case.} A study on the model sensitivity to this hyperparameter is presented in Section~\ref{section:sensitivity_analysis}.

\subsubsection{Implicit regularization of the posterior}\label{sec:regularization_posterior}

Note that the proposed RBC framework relies in probabilistic classifiers or generative models. Although many deep learning classifiers are currently trained based on the cross-entropy loss, which leads to a maximum likelihood estimation of the class labels~\citep{barberBRML2011}, very flexible models, such as deep neural networks, can lead to overconfident classification results, i.e., there being some $j$ such that $p(C_t=j|\bz_t)\approx 1$.
This can be damaging when such models are integrated into the proposed RBC framework since such overconfidence diminishes the relevance of the prior information obtained in previous time instants through the recursion. To remedy this issue, we propose to empirically reduce the confidence in the predictions of deep learning models before integrating them into the proposed framework, with this simple relation
\begin{equation}
    \label{eq:normalization}
    p\left(C_t|\bz_t\right) = \frac{p^\prime\left(C_t|\bz_t\right) + \lambda}{\sum_{C_t'\in\mathscr{C}}\left(p^\prime\left(C_t'|\bz_t\right) + \lambda\right)}\;,
\end{equation}
where $p^\prime\left(C_t|\bz_t\right)$ is the distribution of the overconfident discriminative model, being it the probability of the labels $C_t\in\mathscr{C}$ given the pixel value at time instant $t$, and $\lambda\in\amsmathbb{R}_+$ is a positive constant used to slightly push the predicted class probabilities towards $\frac1{K}$ (i.e., towards a discrete uniform distribution). The same idea can be applied to an overconfident generative model $p^\prime(\bz_{t}|C_t)$~\citep[Chapter~20.3]{barberBRML2011}.
%

The proposed approach is motivated, at a high level, from the \textit{maximum entropy principle}~\citep{jaynes_infotheorystatisticalmechanics}. This principle states that among all available solutions that fit some measurements, the most suitable solution is the one with the highest entropy. In this paper, an estimation problem has not been defined for this matter because the PMFs (i.e., the likelihood or class posterior) are directly obtained. Consequently, the PMF entropy is increased in an \textit{ad-hoc} fashion by following~\refequation{eq:normalization}. This regularization of the posterior, or likelihood, is of great importance within the context of recursive classification of time-series data since overconfident classifiers can mask the prior information from previous time instants, jeopardizing the algorithm performance.
%


\subsubsection{Computational Overhead of the Recursion}
\credreview{
Neural networks or more complex algorithms that classify entire batches of data often experience a substantial increase in complexity with sequence length~\citep{6841049, Kenduiywo2017}.
In contrast, and given the nature of Bayesian recursion described at the beginning of~\refsection{sec:algorithm}, the computational cost of RBC is not affected by the baseline classifier it is built upon and remains constant for each time step regardless of the length of the image time-series. 
To quantify this, we analytically evaluate the overhead introduced by recursion in terms of operations.
}

\credreview{
Assuming that each operation consists of a sum and product or a sum and division, and given that $p(z_t|C_t)$, $p(C_{t-1}|z_{1:t-1})$ and $p(C_t|z_t)$ are already computed (hence, their evaluation cost is the same as for the instantaneous classifier), we can present the following calculations.
For the RBGM, the overhead cost is $K \times (K^2 + K + 2) = K^3 + K^2 + 2K$ operations. This breaks down into $K$ operations in the summation from the numerator, $K$ operations in the inner summation for the denominator multiplied by $K$ operations in the outer summation also in the denominator, and 2 operations from the division and multiplication by the instantaneous classifier result. This result is then multiplied by $K$ because the expression in~\refequation{eq:methods:posterior_generative_model_RBGM} needs to be computed for each class.
By applying the same logic, the overhead cost for the RBDM can be expressed as $K\times(K\times(K+1)+K+2)=K^3+2K^2+2K$.
We can argue that this is not substantial in the context of most classifiers since $K$ is generally small.
For detailed quantitative results on the time required by recursion and instantaneous classifiers across all test sites and algorithms tested in this study, please refer to~\refsection{sec:computational_cost_analysis}.
}

\subsection{Spectral Index Classification (SIC)}
\label{section:recursive_spectral_index_classification}
\begin{figure}[h]
    \centering
    \includegraphics[width=17cm]{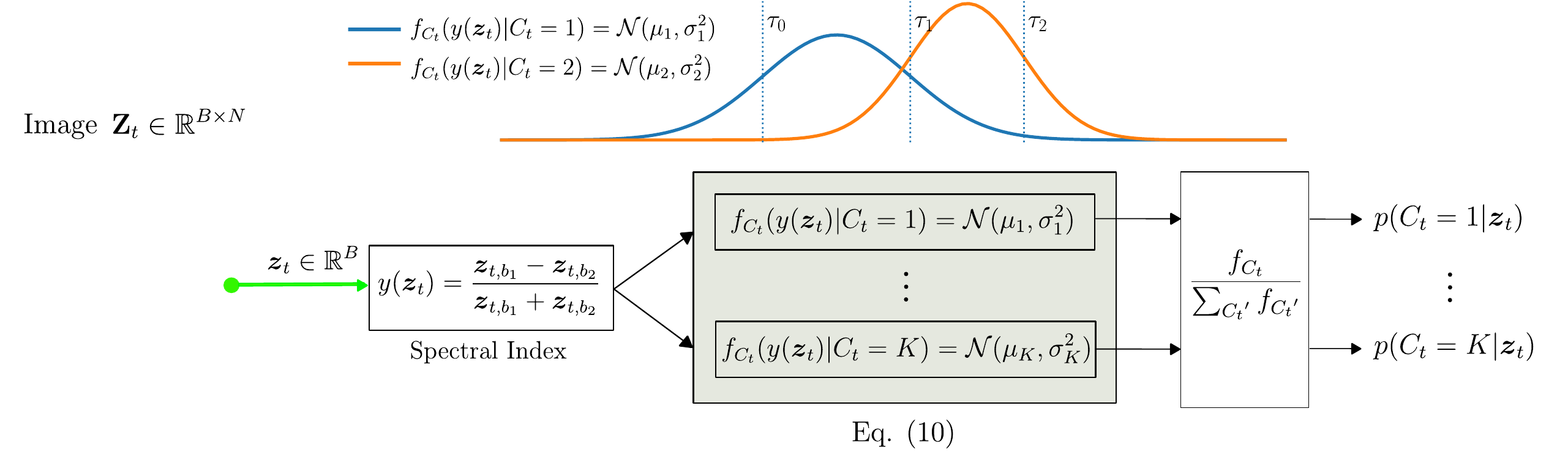}
    \caption{\cred{Overview of the proposed Spectral Index Classifier (SIC),  where the values of standard broadband spectral indices such as the NDWI are converted into probabilities $ p(C_t|\bz_t)$ as in~\refequation{eq:methods:scaled_index_model_conditional_PMF}. The proposed RBC framework can be applied on top of the SIC algorithm, resulting in the recursive SIC (RSIC) model.}}
    \label{fig:sic_diagram}
\end{figure}

We introduce the SIC algorithm, which uses broadband spectral indices to generate the predictive probability of occurrence of land classes, such as water or soil. 
Spectral indices are of interest for classification algorithms due to their clear interpretability and lack of supervision, as explained in~\refsection{sec:intro:background}. 
%
\cred{An overview of this classifier can be found in~\reffigure{fig:sic_diagram}.} The class probability $p(C_t|\bz_t)$ is defined as
\begin{equation}
    \label{eq:methods:scaled_index_model_conditional_PMF}
    p(C_t|\bz_t) = 
    \frac{f_{C_t}\left(y(\bz_t)\right)}{\sum_{C_t' \in \mathscr{C}}f_{C_t'}(y(\bz_t))}\,,
\end{equation}
where $y(\bz_t)$ corresponds to the spectral index value, which is computed as a function of the pixel $\bz_t$. This is a similar but not equivalent idea to applying a softmax function. To compute the probability value, we use a Gaussian function as $f_{C_t} = \mathcal{N}\left(\mu_{C_t},\sigma_{C_t}\right)$. A different value of mean and standard deviation can be assigned for each class as $\boldsymbol{\mu} = \text{col}\{\mu_{C_t}\}$ and $\boldsymbol{\sigma} = \text{col}\{\sigma_{C_t}\}$, being $\text{col}\{ \cdot \}$ the operator returning a vector whose elements are $\mu_{C_t}$ and $\sigma_{C_t}$ for $C_t\in\mathscr{C}$, respectively. The function $f_{C_t}$ can be expressed as
%
\begin{equation}
\label{eq:sic}
    f_{C_t}\left(y(\bz_t)\right) =
    \frac1{\sigma_{C_t}\sqrt{2\pi}}
    \exp\left(-\frac1{2}\left(\frac{y(\bz_t)-\mu_{C_t}}{\sigma_{C_t}}\right)^{\!2}\right)\,,
\end{equation}
%
thus giving a measure of how close the spectral index $y(\bz_t)$ is to the mean value of each class $C_t$, denoted as $\mu_{C_t}$. This is used as an indication of the likelihood of $\bz_t$ being of class $C_t$. The standard deviation $\sigma_{C_t}$ accounts for the length of the spectral index interval defining class $C_t$. For ease of exposition, let us consider for the remainder of this section that $C_t\in\mathscr{C}=\{1,\ldots,K\}$, and also that the class indices are ordered in the same way as the threshold intervals, i.e., class $i$ corresponds to the $i$-th spectral index interval.

The length of the intervals defining each class can be highly non-homogeneous and depends on the spectral index class thresholds $\tau_i$, where $i\in\{0,\hdots,K\}$. These thresholds define a hard classification result based on the spectral index value, with pixel $\bz_t$ being assigned to the $i$-th class if and only if $y(\bz_t)\in(\tau_{i-1},\tau_{i}]$. Their length can be calculated as $L_j = \tau_j - \tau_{j-1}$, where $j\in\{1,\hdots,K\}$. The values of $\boldsymbol{\mu}$ and $\boldsymbol{\sigma}$ are calculated as $\mu_j = L_j/2 + \tau_{j-1}$ and $\sigma_j = L_j/2$, so that the probability of a pixel belonging to a given class decreases smoothly as $y(\bz_t)$ moves away from the center of the interval and approaches one of the thresholds.
The threshold values are determined empirically and are experiment-dependent, as discussed in~\refsection{sec:evaluation}.
%

%

\subsection{Experimental setup}
\label{section:experimental_setup}
A classifier based on a GMM, an LR classifier, and the SIC algorithm introduced in~\refsection{section:recursive_spectral_index_classification}, are compared to their recursive counterparts, namely the RGMM, RLR and RSIC algorithms. 
\cred{When working with data from test sites 1a, 1b, and 2, two additional pre-trained deep learning models are used as a benchmark in the context of water mapping:} the DeepWaterMap~\citep{8013683} and the WatNet~\citep{luo2021_watnet} algorithms.
All models under consideration are listed in Table \ref{tab:algorithms}.
To ensure consistency when evaluating the RBC framework in different areas of study, we maintain the same number of classes across the three test sites. 

%
%
%
%
As discussed in the introduction, the instantaneous likelihood or class posterior to which the RBC framework is applied may be either semi-supervised, supervised or unsupervised. When the data model is unsupervised, the entire procedure may be viewed as unsupervised. The converse is also true for supervised methods. When we describe the methodology followed in the \textit{training stage} in~\refsection{sec:training}, we are referring to the training stage of the GMM and LR models, which are the models that need supervision.
%

\subsubsection{Data splitting}\label{sec:data_split}
\begin{figure}[ht]

  \begin{subfigure}{\linewidth}
  \centering
    \includegraphics[width=16cm]{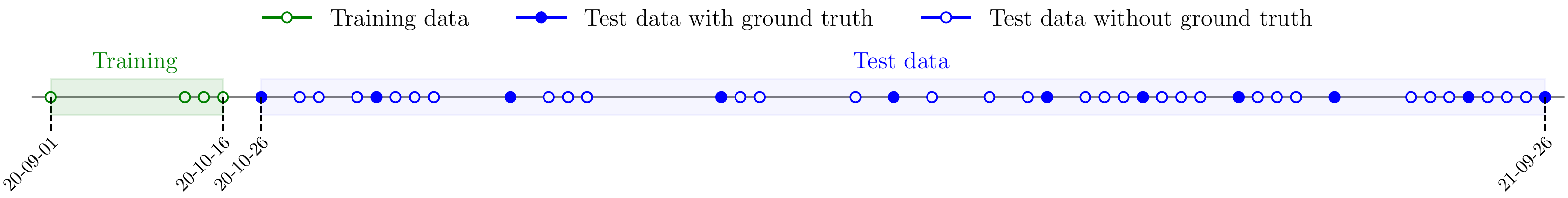}
    \vspace*{-0.7cm}
    \caption{Test sites 1a and 1b}
  \end{subfigure}  

\vspace*{0.4cm}
  \begin{subfigure}{\linewidth}
  \centering
    \includegraphics[width=16cm]{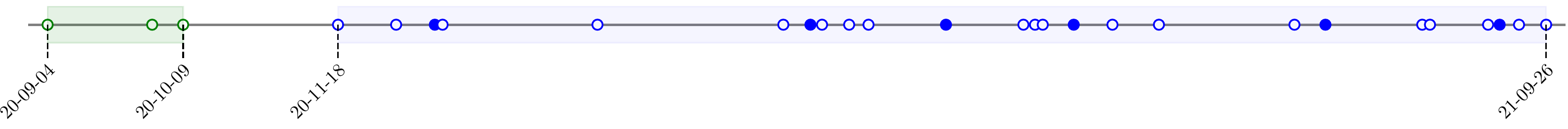}
    \vspace*{-0.6cm}
    \caption{Test site 2}
  \end{subfigure}  
  
\vspace*{0.5cm}
    \begin{subfigure}{\linewidth}
  \centering
    \includegraphics[width=16cm]{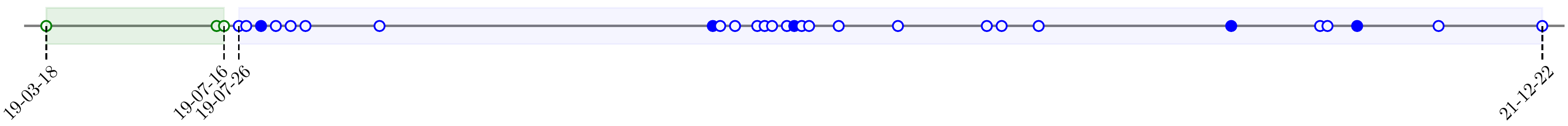}
    \vspace*{-0.65cm}
    \caption{Test site 3}
  \end{subfigure}
  \caption{\cred{Proportionally scaled timeline showing Sentinel-2 image dates \credreview{(yy-mm-dd)} used to train the GMM and LR models (green markers), and to test the GMM, LR, SIC, DWM and WN models, and their recursive counterparts (blue markers). The filled blue markers correspond to test images for which manually generated ground truth labels are available. This allows to evaluate the proposed framework quantitatively.}}  
  \label{fig:timeline}
\end{figure}

\cred{A proportionally scaled timeline with the dates of images used for training and evaluation can be found in~\reffigure{fig:timeline}.}
%
%
%
Each downloaded image belongs to a different date and they are mostly spaced 5 days apart, i.e., the temporal resolution of Sentinel-2 satellites. However, temporal spacing between images may vary as a consequence of filtering images with high cloud/shadow cover and other discrepancies. On the one hand, large temporal spacings between training images translate into training data diversity. On the other hand, large temporal spacings between evaluation images can pose a challenge, because a change in land that occurs gradually can be interpreted as a sudden artifact to be discarded by the recursive algorithm.
\cred{This matter is further discussed in~\refsection{section:discussion}}.
%

%
\subsubsection{Training the GMM and LR models}\label{sec:training}
%

%
%

%
%
The LR and GMM models are trained in a weakly supervised approach. \cred{Training images correspond to those acquired on the dates indicated with green markers in~\reffigure{fig:timeline}.} To generate surrogate ground truth class labels, or pseudo-labels, the pixels from the training images are classified based on their spectral index value (MNDWI for test sites 1a, 1b and 2, and NDWI for test site 3) and considering the class thresholds $\boldsymbol{\tau}$ in~\reftable{tab:parameter_settings}. 
%
%
%
To obtain the generative model~$p(z_t|C_t)$ used in the RBGM from~\refequation{eq:methods:posterior_generative_model_RBGM}, one GMM is trained for each class label, i.e., $p(z_t|C_t)$ is a GMM for each choice of $C_t$. To adequately represent the training pixels without overfitting, we select the smallest number of components for each GMM such that the histograms of the training data distribution and the one generated by the respective GMM are visually close.

\subsubsection{Evaluation}\label{sec:evaluation}

\cred{Test images correspond to those acquired on the dates indicated with blue markers in~\reffigure{fig:timeline}.}
\cred{Across all test sites, the dataset is imbalanced, with the majority of pixels being attributed to the \textit{land} class in the two water mapping experiments and the \textit{forest} class in the deforestation detection experiment. To prevent biases and ensure equal contribution from each class when benchmarking between classification models, balanced classification accuracy is used as a metric for the comparative analysis between instantaneous classifiers and their recursive counterparts in~\refsection{section:error_classification_maps} and~\refsection{sec:quantitative_analysis}.}
\cred{In Section \ref{section:sensitivity_analysis}, the same metric is used to assess the model sensitivity to the transition probability hyperparameter.}

\textbf{Ground truth:} The lack of openly accessible labeled Sentinel-2 data for time-series analysis presents a significant challenge to the assessment of our framework.
\cred{Consequently, water labels were manually generated for the dates employed in the quantitative analysis (filled blue markers in~\reffigure{fig:timeline}), and shared by the authors at~\citep{zenodo_link}. This procedure was facilitated by the LabelStudio tool~\footnote{\url{https://labelstud.io/}}.}
In the case of test site 3, the MultiEarth challenge dataset contains deforestation labels, \cred{which removes the need to manually generate ground truth labels for the third experiment.} The dates for which deforestation labels are available do not necessarily match any date from the Sentinel-2 images in the MultiEarth dataset. Taking this into account, for each of the provided five labels, error classification maps are computed between the label and the classification result that is closest in time after the label date.

\begin{figure}[h]
    \centering
    \includegraphics[width=10cm]{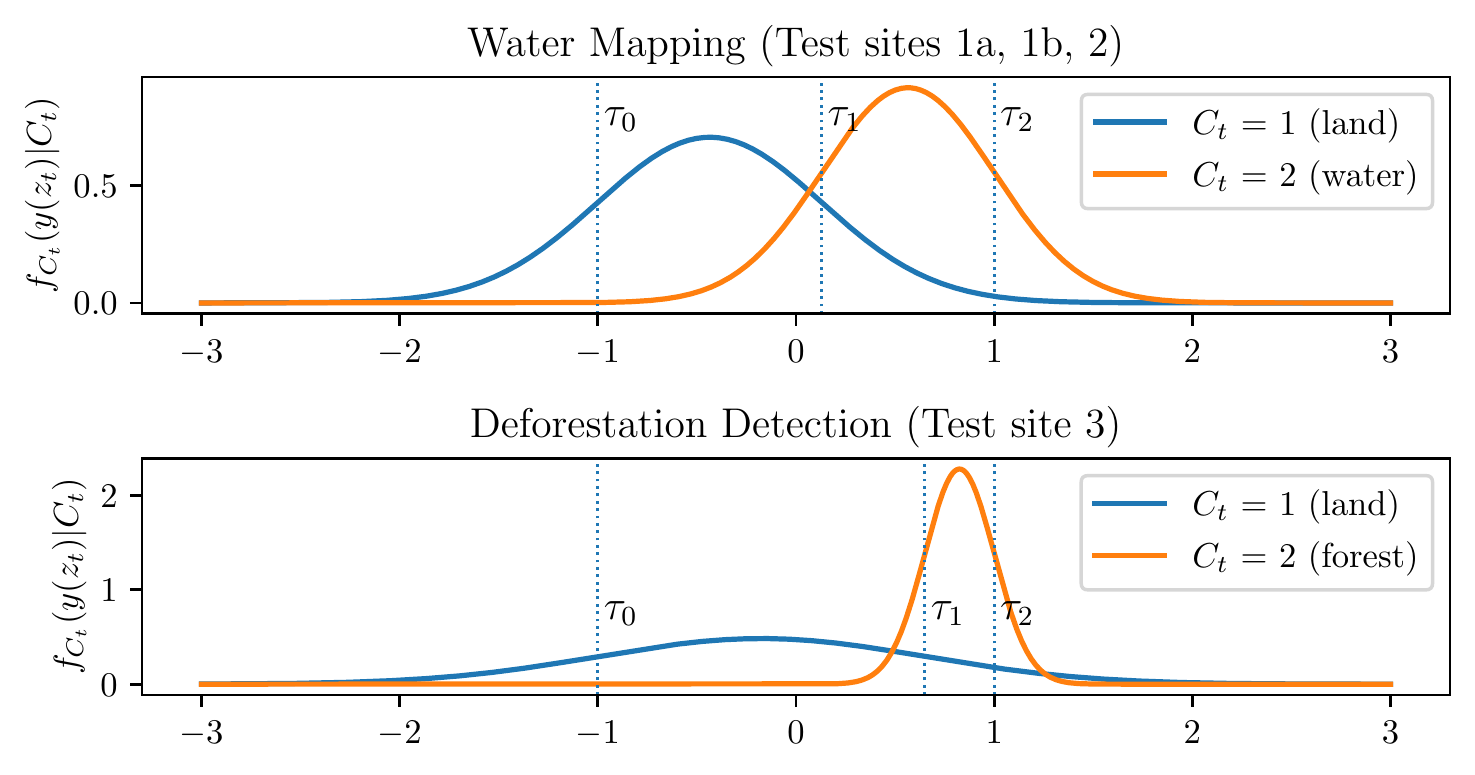}
    \caption{\cred{Function $f_{C_t}\left(y(\bz_t)\right)$ introduced in~\refsection{section:recursive_spectral_index_classification}. The corresponding values of $\boldsymbol{\tau}$, $\boldsymbol{\mu}$, $\boldsymbol{\sigma}$, and the spectral index $y(\bz_t)$ are the ones in~\reftable{tab:parameter_settings} for the water mapping experiments (top subplot) and the deforestation detection experiment (bottom subplot).}}
    \label{fig:gaussians_sic}
\end{figure}

\textbf{Parameter settings:}
The parameter values used for the conducted experiments are presented in~\reftable{tab:parameter_settings}.
For the recursive algorithms, $\epsilon$ is chosen to maximize the average balanced accuracy across test images with available ground truth, with optimization done independently for each algorithm and test site. Optimal $\epsilon$ values are determined through sensitivity analysis as described in~\refsection{section:sensitivity_analysis}.
%
%
To prevent the adverse effects of overconfident predictions of the instantaneous classifiers, we empirically selected $\lambda=0.8$ for all methods.
To convert a standard broadband spectral index into a probability measure as in~\refequation{eq:methods:scaled_index_model_conditional_PMF}, it is necessary to define the thresholds $\boldsymbol{\tau}$. These are tuned accordingly so that the generated classification maps obtained with the training images are visually close to the reference maps in~\refmultiplefiguresthree{fig:data_aos:areas_of_study:oroville_dam}{fig:data_aos:areas_of_study:charles_river}{fig:data_aos:areas_of_study:amazon}. As explained in~\refsection{section:recursive_spectral_index_classification}, the thresholds are used to calculate $\boldsymbol{\mu}$ and $\boldsymbol{\sigma}$ (see \reffigure{fig:gaussians_sic}). 
%

{\renewcommand{\arraystretch}{1.5}%
\begin{table}[h]
\small
\centering
    \caption{Parameter settings for the three experiments conducted in this research. The transition probability $\epsilon$ is defined in~\refsection{sec:class_transition_probabilities} and the regularization constant $\lambda$ is introduced in~\refsection{sec:regularization_posterior},~\refequation{eq:normalization}. The thresholds $\boldsymbol{\tau}_\mathrm{W}$ and $\boldsymbol{\tau}_\mathrm{D}$ are used to generate surrogate ground truth labels to train the GMM and LR algorithms, as explained in~\refsection{sec:training}, and to generate predictive probabilities of occurrence with the SIC algorithm (see~\refsection{section:recursive_spectral_index_classification}).}\label{tab:parameter_settings}
\begin{tabular}{|l|ll|}
\hline
\multirow{7}{*}{Test sites 1a, 1b} 

& 
\multicolumn{2}{l|}{$C_t \in \mathscr{C}_\mathrm{W}=\{\mathsf{land},\ \mathsf{water}\}$}\\ 

\cline{2-3}&\multicolumn{1}{l|}{\multirow{3}{*}{SIC}}& 
$\boldsymbol{\tau}_\mathrm{W} = [-1,\ 0.13,\ 1]$;\\
&\multicolumn{1}{l|}{\multirow{2}{*}{}} & $\boldsymbol{\mu}_\mathrm{W}=[-0.435,\ 0.565]$;\\
&\multicolumn{1}{l|}{\multirow{2}{*}{}} &$\boldsymbol{\sigma}_\mathrm{W}=[0.565,\ 0.435]$;\\
&\multicolumn{1}{l|}{\multirow{2}{*}{}} &$y_{\rm{MNDWI}}\left(\bz_t\right) = \frac{z_{t,\rm{green}} - z_{t,\rm{SWIR}}}{z_{t,\rm{green}} + z_{t,\rm{SWIR}}}$\\  

\cline{2-3} 
                               & \multicolumn{1}{l|}{RSIC}                      & $\epsilon$ = 0.001 (1a), 0.02 (1b); $\lambda=0.8 $    \\ \cline{2-3} 
                               & \multicolumn{1}{l|}{RGMM}                      & $\epsilon$ = 0.2, 0.09; $\lambda=0.8  $         \\ 
                             \cline{2-3} 
                               & \multicolumn{1}{l|}{RLR}                      & $\epsilon$ = 0.001, 0.02; $\lambda=0.8  $   \\  
                               \cline{2-3} 
                               & \multicolumn{1}{l|}{RDWM}                      & $\epsilon$ = 0.001, 0.095; $\lambda=0.8  $   \\
                                \cline{2-3} 
                               & \multicolumn{1}{l|}{RWN}                      & $\epsilon$ = 0.005, 0.085; $\lambda=0.8  $   \\\hline
\multirow{6}{*}{Test site 2} 

& 
\multicolumn{2}{l|}{

$C_t \in \mathscr{C}_\mathrm{W}$ 
}                          \\ 
\cline{2-3} 
                               
& \multicolumn{1}{l|}{\multirow{2}{*}{SIC}} 

& 
$\boldsymbol{\tau}_\mathrm{W}$; $\boldsymbol{\mu}_\mathrm{W}$;
$\boldsymbol{\sigma}_\mathrm{W}$\\

& \multicolumn{1}{l|}{\multirow{2}{*}{}} &$y_{\rm{MNDWI}}\left(\bz_t\right)$\\  
\cline{2-3}

                               & \multicolumn{1}{l|}{RSIC}                      & $\epsilon=0.1$; $\lambda=0.8  $    \\ \cline{2-3} 
                               & \multicolumn{1}{l|}{RGMM}                      & $\epsilon=0.001$; $\lambda=0.8  $         \\ 
                             \cline{2-3} 
                               & \multicolumn{1}{l|}{RLR}                      & $\epsilon=0.005$; $\lambda=0.8  $   \\ 
                                \cline{2-3} 
                               & \multicolumn{1}{l|}{RDWM}                      & $\epsilon=0.001$; $\lambda=0.8  $   \\
                                \cline{2-3} 
                               & \multicolumn{1}{l|}{RWN}                      & $\epsilon=0.001$; $\lambda=0.8  $   \\\hline
\multirow{5}{*}{Test site 3} 

& 
\multicolumn{2}{l|}{$C_t \in \mathscr{C}_\mathrm{D}=\{\mathsf{land},\ \mathsf{forest}\}$}\\ 
\cline{2-3}& \multicolumn{1}{l|}{\multirow{2}{*}{SIC}}& 
$\boldsymbol{\tau}_\mathrm{D} = [-1,\ 0.65,\ 1]$;\\
& \multicolumn{1}{l|}{\multirow{2}{*}{}} &
$\boldsymbol{\mu}_\mathrm{D}=[-0.175,\ 0.825]$;\\
& \multicolumn{1}{l|}{\multirow{2}{*}{}} & $\boldsymbol{\sigma}_\mathrm{D}=[0.825,\ 0.175]$\\
& \multicolumn{1}{l|}{\multirow{2}{*}{}} & $y_{\rm{NDWI}}\left(\bz_t\right) = \frac{z_{t,\rm{green}} - z_{t,\rm{NIR}}}{z_{t,\rm{green}} + z_{t,\rm{NIR}}}$ \\  

\cline{2-3} 
                               & \multicolumn{1}{l|}{RSIC}                      & $\epsilon=0.03$; $\lambda=0.8  $    \\ \cline{2-3} 
                               & \multicolumn{1}{l|}{RGMM}                      & $\epsilon=0.04$; $\lambda=0.8  $         \\ 
                             \cline{2-3} 
                               & \multicolumn{1}{l|}{RLR}                      & $\epsilon=0.04$; $\lambda=0.8  $   \\   \hline
\end{tabular}
\end{table}
{\renewcommand{\arraystretch}{1.1}%


{\renewcommand{\arraystretch}{1.1}%

\section{Results}
\label{section:results}

%
Complete results can be reproduced following the instructions in~\url{https://github.com/neu-spiral/RBC-SatImg}.
%
%
\cred{Overall, the RBC framework significantly increases the robustness of existing classification algorithms in multitemporal settings, while preserving the adaptability to changes in the land. Substantial improvements are shown when compared to pre-trained state-of-the-art deep learning-based classifiers, without the need for additional training data.}
%
%


%
%
\begin{table}[h]
    \centering
    \caption{Full name and abbreviation of the benchmark classification models, i.e., SIC (introduced in~\refsection{section:recursive_spectral_index_classification}), GMM, LR, DWM~\citep{8013683} and WN~\citep{luo2021_watnet}, and their recursive counterparts, i.e., RSIC, RGMM, RLR, RDWM and RWN, which derive from applying the proposed recursive Bayesian classification framework.} 
    \begin{tabular}{l c}
        \hline
        Full Name & Abbreviation \\ \hline 
        Spectral Index Classifier (\refsection{section:recursive_spectral_index_classification}) & SIC \\
        Gaussian Mixture Model & GMM \\
        Logistic Regression & LR \\
                DeepWaterMap~\citep{8013683}  & DWM\\
        WatNet~\citep{luo2021_watnet} & WN \\
        Recursive Spectral Index Classificatier & RSIC\\
        Recursive Gaussian Mixture Model & RGMM \\
        Recursive Logistic Regression & RLR\\
         Recursive DeepWaterMap & RDWM  \\
        Recursive WatNet & RWN\\
        \hline
    \end{tabular}
    \label{tab:algorithms}
\end{table}

\subsection{Error classification maps}
\label{section:error_classification_maps}
\cred{The following figures include the classification maps, error classification maps, and balanced classification accuracy results obtained using the test images with available ground truth labels. RGB composite images of the studied areas are shown as a reference because they highlight changes in the scene. In the case of test site 1b, and due to space limitations, only one every two test images with ground truth data are included in the analysis. The interested reader may find the classification map results for all test images, including those without ground truth data, in the supplemental material.}

\subsubsection{Test site 1}

Water mapping results for test sites 1a and 1b are presented in~\refmultiplefigures{fig:results:water_mapping_OD_downstream}{fig:results:water_mapping_OD_upstream}, respectively. 
%
%
\reffigure{fig:results:water_mapping_OD_downstream} shows noticeable differences among benchmark and recursive algorithms. For instance, the SIC algorithm classifies a large portion of the stream as land for dates 2021-05-19, 2021-06-13, 2021-09-06 and 2021-09-26. The same is observed with the LR and DWM classifiers for dates 2021-05-19 and 2021-09-06. Nevertheless, their recursive counterparts, i.e., the RSIC, RLR and RDWM algorithms, can adequately classify most of the stream pixels as water. \cred{This translates into an increase in balanced classification accuracy of more than 20\% provided by the use of recursion. For instance, the RSIC algorithm provides an improvement of 26.95\% and 20.31\% for dates 2021-05-19 and 2021-06-13.}
The WN algorithm classifies some portions of land as water on dates between 2020-11-25 and 2021-07-08. While the RWN algorithm misclassifies part of the stream, it shows considerably more accurate classification maps. This can be especially observed for the dates 2021-04-09 and 2021-05-19, where the RBC framework offers an improvement in balanced classification accuracy of 8.85\% and 4.9\%, respectively. 
Overall, results suggest that the proposed RBC framework improves the performance of modern deep learning-based mapping algorithms. Moreover, it has been observed that the DWM and WN algorithms provide overconfident classification results, which can be an issue in recursive multitemporal classification. This is solved with the strategy proposed in~\refequation{eq:normalization} to regularize the predictive class posterior.

\begin{figure}[h!]
    \centering
    \includegraphics[width=14cm]{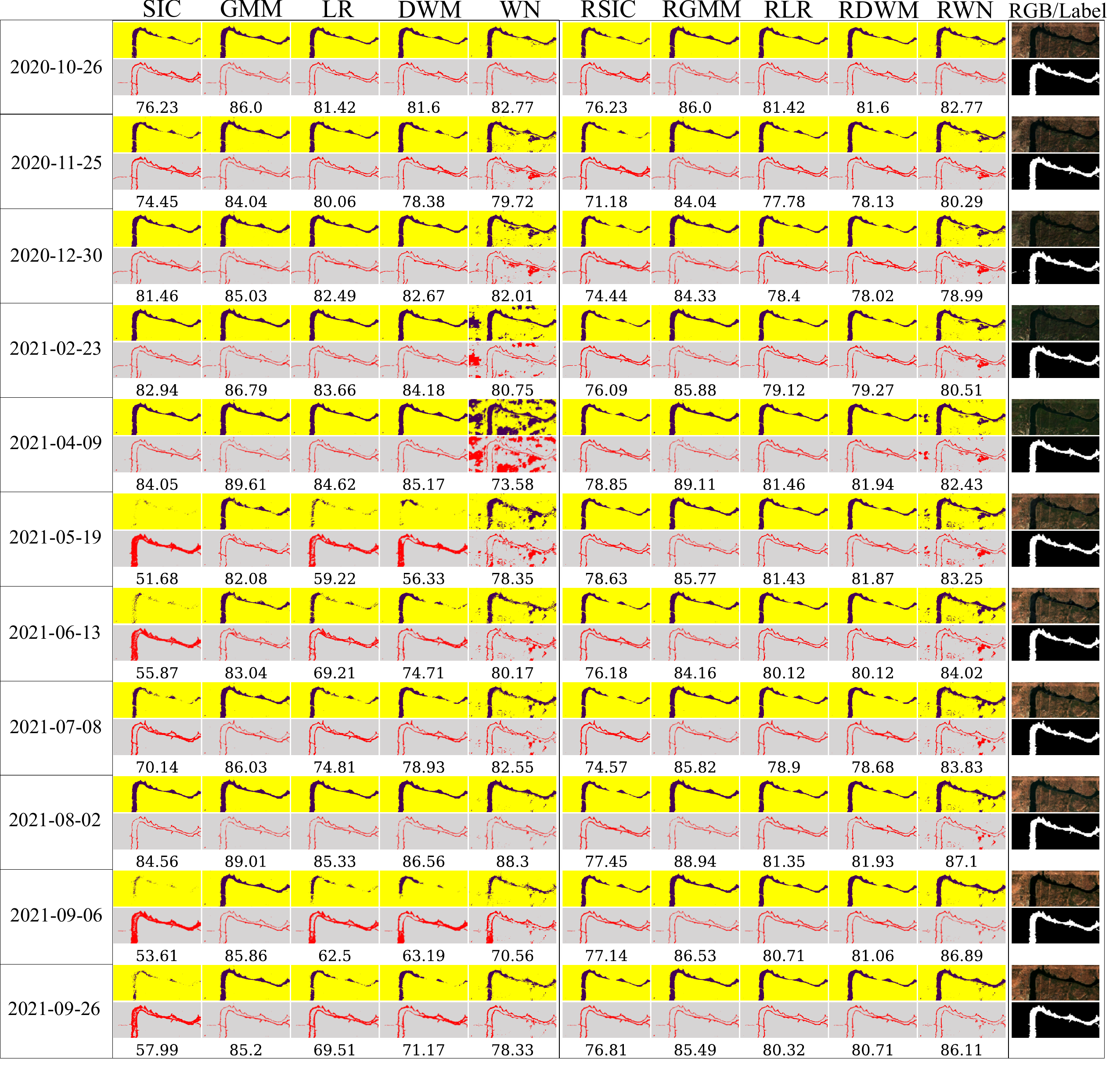}
    \caption{\cred{Water mapping results obtained for the Oroville dam downstream (test site 1a). Classification maps, error classification maps, and balanced classification accuracy results are presented. Purple and yellow represent water and land, respectively.}}
    \label{fig:results:water_mapping_OD_downstream}
\end{figure}

Results in~\reffigure{fig:results:water_mapping_OD_upstream} show that both the instantaneous classifiers and their recursive counterparts are able to adequately capture the decrease in water levels over time in the dam upstream in test site 1b.
In the case of date 2020-12-30, the SIC, LR, DWM and WN algorithms misclassify a considerable amount of water as land, while their recursive counterparts output more correct classification maps. \cred{This results in an increase of balanced classification accuracy of 14.82\%, 4.27\%, 13.81\%, 9.62\%, and 11.03\% for the RSIC, RGMM, RLR, RDWM and RWN algorithms respectively. Between these, the improvement provided by the RGMM classifier is more modest due to its non-recursive counterpart already performing well. A similar effect is observed on dates such as 2021-06-13, 2021-08-02, and 2021-09-26, for which the non-recursive methods provide adequate classification results and consequently the improvement introduced by recursion is moderate (under 10\%).} 
\cred{The trade-off between adaptability and robustness presents a challenge to the recursive framework for the upstream subscene of Oroville Dam. Prioritizing robustness makes the recursive framework less flexible, potentially leading to delayed detection of abrupt scene changes. For example, the increase in water level starting in February is better identified by the instantaneous classifiers, as they do not rely on information from previous images showing lower water levels. Consequently, when recursion is employed on the date 2021-04-09, there is a decrease in balanced classification accuracy of up to 6.79\%, as indicated by the RSIC classifier. However, this setback is gradually resolved on subsequent dates, as indicated by improved classification results on 2021-06-13.}
%

%

%
\begin{figure}[h!]
    \centering
    \includegraphics[scale=0.43]{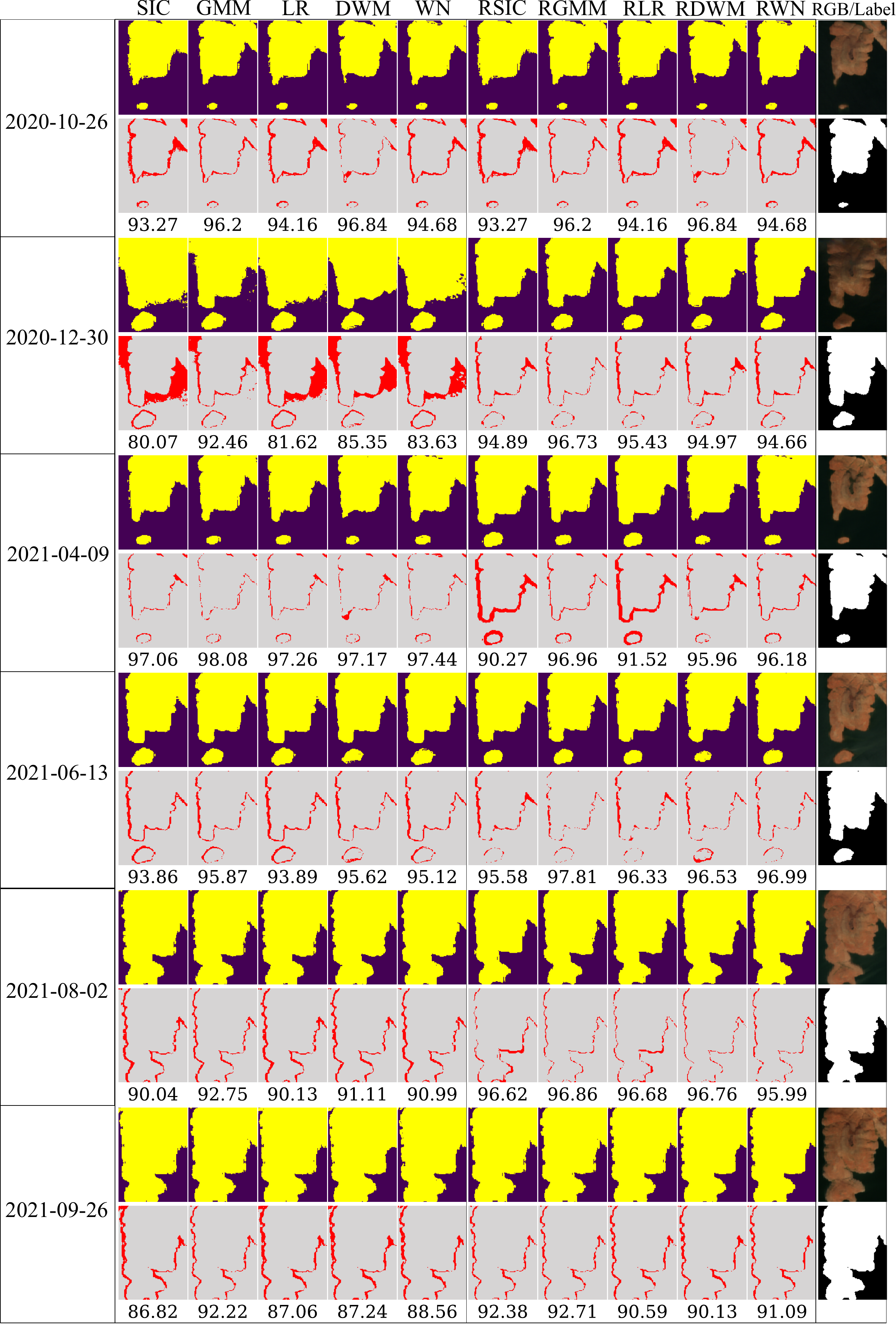}
    \caption{\cred{Water mapping results obtained for the Oroville dam upstream (test site 1b). Classification maps, error classification maps, and balanced classification accuracy results are presented. Purple and yellow represent water and land, respectively.}}
    \label{fig:results:water_mapping_OD_upstream}
\end{figure}

\subsubsection{Test site 2}

Water mapping results for test site 2 are presented in~\reffigure{fig:results:Site_c_results_2class}. \cred{The interested reader may refer to~\refappendix{appendix:charles_river}, where classification map results are shown for a three-class classification experiment with the same data, thus demonstrating the scalability of the framework to handle more complex classification tasks.}
%
%
Test site 2 extends over an area covering the Boston harbor, the Charles river lower, mid and some upper basins, and the Mystic river lower basin. Since most of these correspond to urban and suburban areas, we can find many reflective surfaces from, e.g., building terraces and metal sheds, which lead to pixels with high spectral reflectance. Such pixels may be easily misclassified as water since their MNDWI values are close to zero. This can be observed in the classification maps from 2020-12-13, where the recursive algorithms result in fewer misclassifications of reflective surface pixels as water. \cred{However, this reduction in misclassifications does not necessarily translate to a performance increase in terms of balanced classification accuracy. In general, results for dates between 2020-12-13 and 2021-04-24 are already good enough for the instantaneous classifiers. Consequently, the improvement provided by their recursive counterparts is either non-existent or very small.}
On 2021-05-27, given the appearance of cyanobacterial blooms, due to which the water becomes diluted with chlorophyll pigments, an important portion of water pixels are classified as land by the instantaneous SIC, GMM, LR and DWM classifiers, whereas their recursive versions provide more accurate classification maps. \cred{This leads to an improvement in balanced classification accuracy of 12.77\%, 12.4\%, 8.59\%, and 9.11\% for the RSIC, RGMM, RLR and RDWM models respectively. The WN algorithm exhibits poor performance, particularly since 2021-05-27, misclassifying a significant portion of land pixels as water. We observe an increase in balanced classification accuracy provided by its recursive counterpart of 10.58\% on 2021-05-27, 6.02\% on 2021-07-31, and 0.95\% on 2021-09-14. The reason for this decrease in improvement over time is due to the repeated failures in the task of the WN instantaneous classifier.} Overall, the recursive algorithms provide significantly more robust performance than their non-recursive counterparts. The latter are less sensitive to atmospheric interference and illumination factors.
For some dates, the recursion introduces a smoothing effect, which makes it more difficult to adapt to class changes. This can be understood by comparing the classification maps obtained with the SIC and RSIC algorithms on 2021-03-20.



\begin{figure}[h!]
\centering
{\includegraphics[width=14cm]{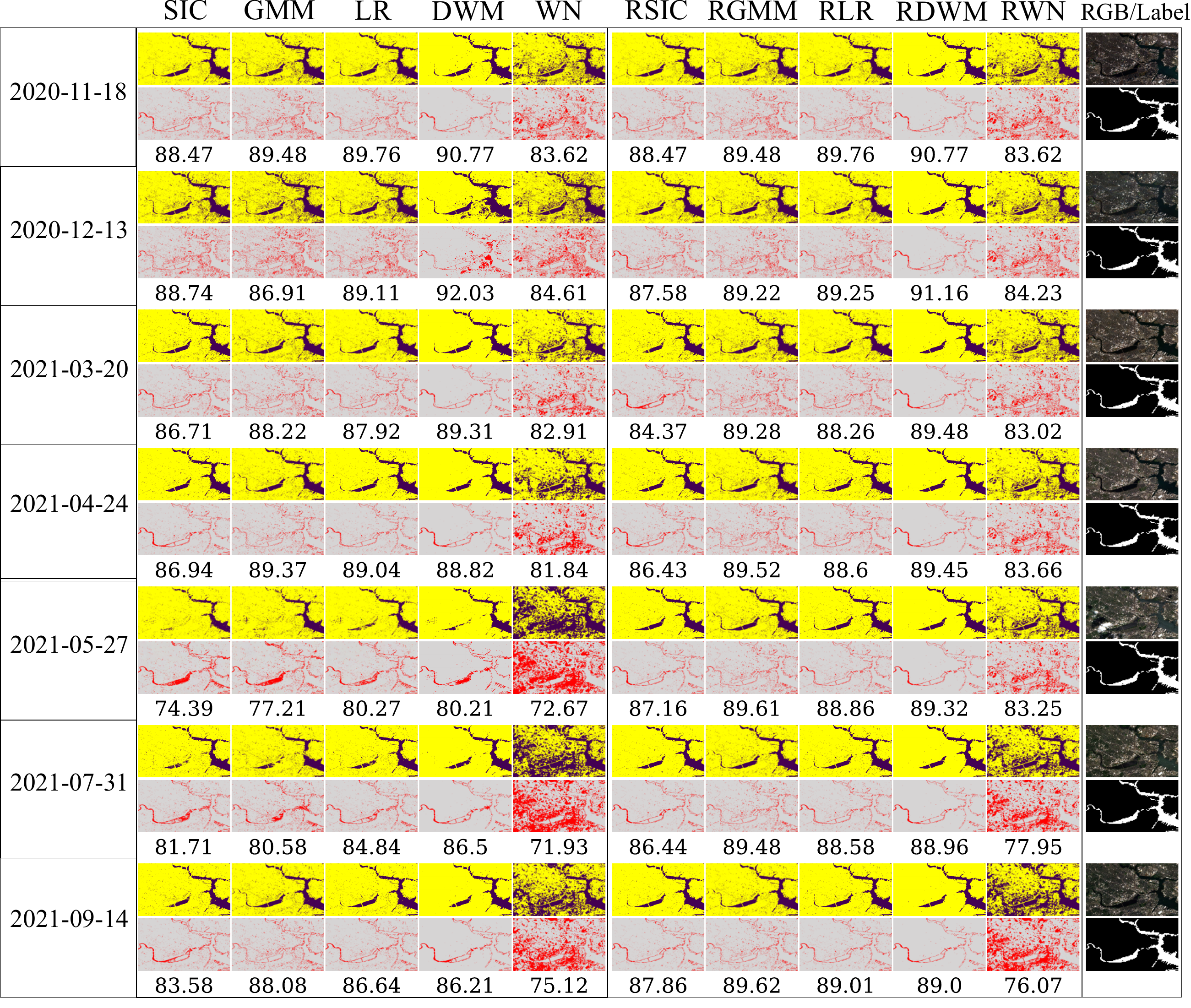}}
\caption{\cred{Water mapping results obtained for the Charles river basin (test site 2). Classification maps, error classification maps, and balanced classification accuracy results are presented. Purple and yellow represent water and land, respectively.}}
\label{fig:results:Site_c_results_2class}
\end{figure}
\subsubsection{Test site 3}
Deforestation detection results for test site 3 are illustrated in~\reffigure{fig:results:experiment_3_qualitative_analysis}. On dates 2020-06-10, 2020-08-04, and 2021-05-26, cloud presence disrupts the performance of the instantaneous classifiers, while their recursive counterparts demonstrate adequate classification map results.
\cred{This led to a rise in balanced classification accuracy of 7.06\%, 14.17\%, and 8.37\% for the RSIC, RGMM, and RLR algorithms on 2020-06-10, and 15.25\%, 10.58\%, and 14.7\% on 2021-05-26. On 2020-08-04, despite improved classification maps, there is a loss in balanced classification accuracy of 1.6\% and 1.88\% for the RGMM and RLR algorithms respectively. We attribute this to the class imbalance in the dataset, where the forested area substantially exceeds the deforested area. The RSIC algorithm, however, provides a slight improvement of 0.97\% for that date.}
The adaptability of the framework is evident in the classification map results from 2021-05-26, which reveal a newly deforested area detected by the three recursive algorithms but missed by their non-recursive counterparts. On the subsequent date (2021-08-19), only the RSIC algorithm can detect this same deforested area, which remains undetected for the other algorithms. This failure of the RGMM and RLR algorithms is due to repeated failures in the task by the instantaneous classifiers, which disrupts the performance of their recursive counterparts.
Supplemental results show that the recursive classifiers need two iterations where the instantaneous classifiers detect the deforested area to acknowledge this change, as a consequence of the trade-off between the robustness and adaptability of the framework.

\begin{figure}[h!]
\centering
{\includegraphics[width=0.65\linewidth]{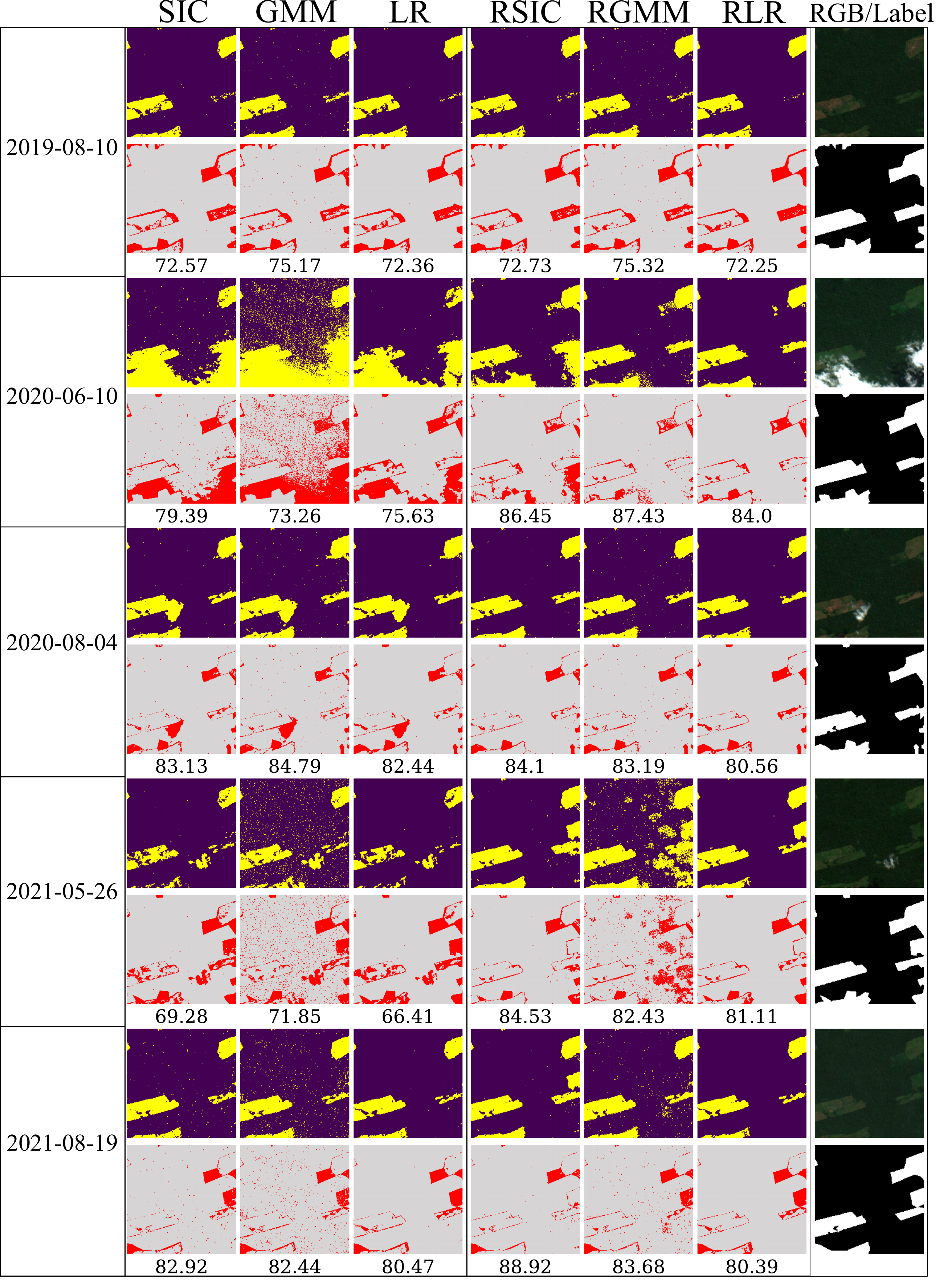}}
\caption{\cred{Deforestation detection results obtained for the Amazon rainforest area (test site 3). Classification maps, error classification maps, and balanced classification accuracy results are presented. Purple and yellow represent forest and deforested area, respectively.} For this experiment, Sentinel-2 images and deforestation labels have been obtained from the MultiEarth challenge dataset~\citep{cha2023multiearth}.}
\label{fig:results:experiment_3_qualitative_analysis}
\end{figure}

\subsection{Classification accuracy visualization}\label{sec:quantitative_analysis}
\cred{The boxplot in~\reffigure{fig:boxplot} shows the distribution of the balanced classification accuracy results presented in the previous subsection.}
\cred{The introduction of recursion mitigates the negative outliers from the non-recursive models. Additionally, negative outliers from the recursive models fall within the interquartile range of the non-recursive models, or are slightly below their lower quartile in the case of the RGMM for test site 1b and the RLR for test site 3. For instance, when analyzing the RSIC model with data from test site 1a, there is a significant reduction in result variability, with one negative outlier falling within the interquartile range of the SIC model.}

\cred{A noticeable trend is the reduced variability in performance among the recursive algorithms, evident from the narrower spread of balanced accuracy values. This is specially apparent for the RGMM model. While the upper quartiles remain consistent and do not show a significant increase, the lower range of results for the recursive models is notably higher than the ones offered by their recursive counterparts. This suggests that although peak accuracies do not exhibit a significant rise, there is a marked enhancement in the lower-end performance, implying a more consistent and improved overall performance across the tested algorithms. This results from the robustness provided by the RBC framework. Upper quartiles for the recursive classifiers for test sites 1b and 3 indicate a noticeable enhancement. Finally, in the case of the RSIC, RLR, and RDWM algorithms when tested with data from test site 1a, we observe a decrease in the upper quartile values that may have originated from the higher variance in the accuracies provided by their instantaneous classifiers.}

\begin{figure}[h!]
\centering
{\includegraphics[width=0.98\linewidth]{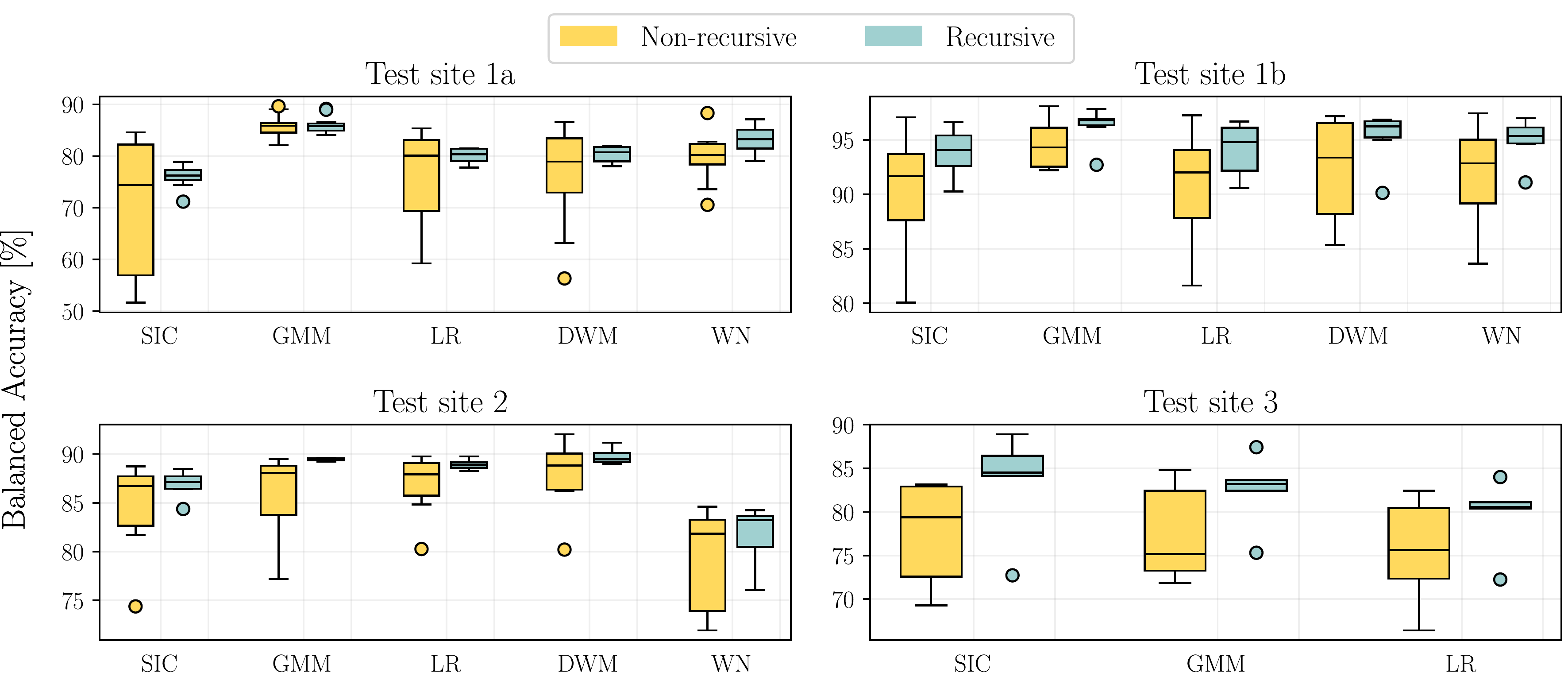}}
\caption{\cred{Boxplot showing the distribution of balanced accuracy results for the SIC, GMM, LR, DWM and WN algorithms (yellow boxes), including their recursive counterparts (blue boxes). The boxplot illustrates the median with a line within each box, the interquartile range with the box boundaries, and the data range with whiskers. Outliers are represented as filled individual circles.}}
\label{fig:boxplot}
\end{figure}

\subsection{Sensitivity analysis}
\label{section:sensitivity_analysis}
The class transition probabilities are governed by the hyperparameter $\epsilon$ introduced in~\refsection{sec:class_transition_probabilities}. This hyperparameter is site-specific, emphasizing the need for careful selection of its value. 
An analysis of the RBC model sensitivity to $\epsilon$ has been conducted using the data with available ground truth for each test site. This provided the optimal values for the hyperparameter, which have been used in the three conducted experiments and can be found in~\reftable{tab:parameter_settings}.
The results in~\reffigure{fig:sensitivity_analysis_1a} show the balanced classification accuracy concerning values of $\epsilon$ between 0.001 and 0.8 for the water mapping experiment with data from test site 1a. 
\reffigure{fig:sensitivity_analysis_multiearth} shows results for data from test site 3 in the context of deforestation detection.  

Given that we consider uniform marginal class probabilities, results for $\epsilon=0.5$ match the non-recursive benchmark given by the instantaneous classifiers. The improvement with respect to the non-recursive algorithms is notable for $\epsilon<0.5$. For the water mapping task, the RBC framework provides an increase in average balanced classification accuracy of up to 5.87\%, 0.3\%, 4.38\%, 3.68\%, and 4.55\% for the SIC, GMM, LR, DWM, and WN classifiers respectively. The RGMM algorithm gives a lower improvement because its non-recursive counterpart provides an already good performance, which can be understood from the low variance of the non-recursive GMM boxplot for test site 1a in~\reffigure{fig:boxplot}. For the deforestation detection task, the improvements in accuracy are of up to 5.9\%, 4.9\%, and 4.2\% for the SIC, GMM, and LR classifiers respectively.
%
%
For $\epsilon>0.5$, the recursive algorithms demonstrate inferior performance compared to their non-recursive counterparts. This makes sense, as $\epsilon>0.5$ suggests a higher likelihood for a pixel to transition between classes rather than remaining in the same class, which is an unrealistic hypothesis. Although the accuracy provided by the RSIC model increases slightly for $\epsilon > 0.6$, this result does not appear to be influential in shaping the conclusions.

The best performance is achieved for $0 < \epsilon < 0.1$. This region is shaded in grey and magnified in the right subplot of the figures. The $\epsilon$ values yielding the best results are 0.001, 0.2, 0.001, 0.001, and 0.005 for the RSIC, RGMM, RLR, RDWM, and RWN models in the water mapping analysis, and 0.03, 0.04, and 0.04 for the RSIC, RGMM, and RLR models in the deforestation detection analysis.
%
The RGMM algorithm in the water mapping analysis shows a wider range of $\epsilon$ values providing classification accuracy results near the best accuracy, and a notably lower sensitivity to $\epsilon$ variations. The sensitivity to this hyperparameter is highly influenced by the performance of the instantaneous classifier, in particular to the variance of its balanced classification accuracy results. As an example, the RSIC algorithm shows the highest sensitivity to $\epsilon$ in~\reffigure{fig:sensitivity_analysis_1a}, being the SIC algorithm the one showing the highest interquartile range in the subplot from~\reffigure{fig:boxplot} for test site 1a. Interquartile ranges in the subplot for test site 3 are similar between the three algorithms under study, resulting in similar sensitivities to $\epsilon$ in~\reffigure{fig:sensitivity_analysis_multiearth}.
%


%

\begin{figure}[h!]
     \begin{subfigure}[b]{1\textwidth}
     \hspace{-1.1cm}
         \includegraphics[width=1.1\linewidth]{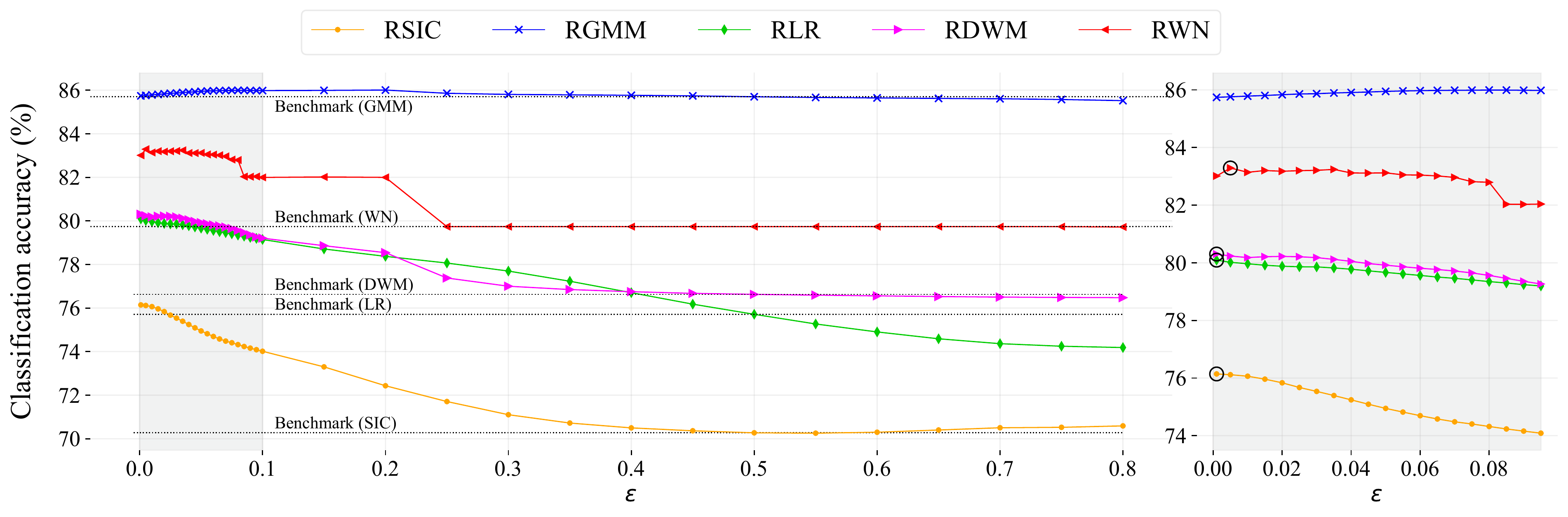}
         \caption{Sensitivity analysis for the water mapping task (test site 1a). Maximum accuracy is given by $\epsilon = 0.001,\ 0.2,\ 0.001,\ 0.001,\ 0.005$ for RSIC, RGMM, RLR, RDWM and RWN.}
         \label{fig:sensitivity_analysis_1a}
     \end{subfigure}
     
     \begin{subfigure}[b]{1\textwidth}
     \hspace{-1.1cm}
         \includegraphics[width=1.1\linewidth]{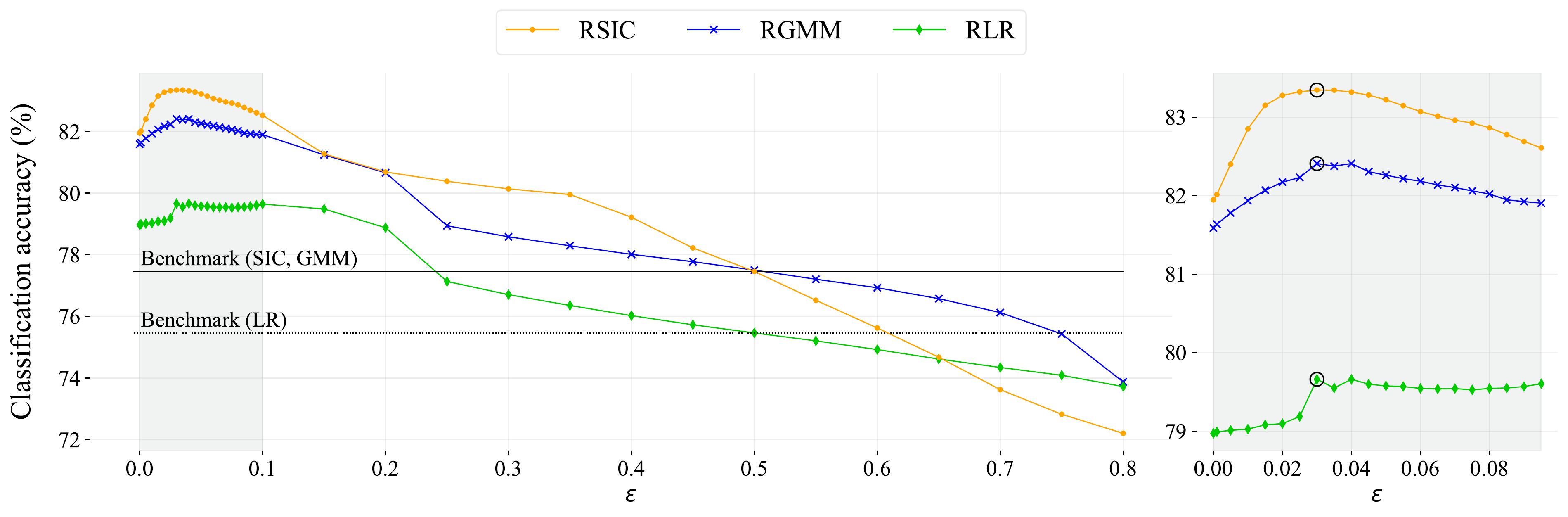}
         \caption{Sensitivity analysis for the deforestation detection task (test site 3). Maximum accuracy is given by $\epsilon = 0.03,\ 0.04,\ 0.04$ for RSIC, RGMM and RLR.}
         \label{fig:sensitivity_analysis_multiearth}
     \end{subfigure}
     \caption{Sensitivity analysis conducted with data from test sites 1a (water mapping) and 3 (deforestation detection). Average balanced classification accuracy results are presented for $\epsilon$ values between 0.001 and 0.8. The best performance is achieved for the region $0 < \epsilon < 0.1$, which is shaded in grey and magnified in the right subplots.}
\end{figure}

\subsection{Computational Cost Analysis}\label{sec:computational_cost_analysis}
\credreview{
We provide an analysis of the computational cost associated with the RBC framework. Specifically, we examine the computation time dedicated to recursion and the time required by the corresponding instantaneous classifier (SIC, GMM, LR, WN, or DWM) for each test site and algorithm, measured over a single time step. RBC transforms any instantaneous classifier into an online method with minimal overhead added by recursion at each time step, as demonstrated by the magnitudes in~\reftable{tab:computational_cost_analysis}.
Using the code in~\url{https://github.com/neu-spiral/RBC-SatImg}, the cost analysis was conducted on an Apple M1 Pro (10 cores, 16-core integrated GPU) with 16 GB RAM, running macOS Sonoma 14.4.1. TensorFlow was used to integrate GPU via the METAL API.}
\begin{table}[h]
\centering
\caption{\textcolor{black}{Total computation time in seconds for recursion and instantaneous classification (baseline) across different test sites for RSIC, RGMM, RLR, RWN, and RDWM algorithms.}}
\arrayrulecolor{black} 
\begin{tabular}{|>{\color{black}}c|>{\color{black}}c|>{\color{black}}c|>{\color{black}}c|>{\color{black}}c|>{\color{black}}c|>{\color{black}}c|}
\hline
\textcolor{black}{Metric} & \textcolor{black}{Test Site} & \textcolor{black}{RSIC} & \textcolor{black}{RGMM} & \textcolor{black}{RLR} & \textcolor{black}{RWN} & \textcolor{black}{RDWM} \\ \hline
\multirow{4}{*}{\centering \textcolor{black}{Recursion Time (s)}} 
                        & 1a &0.008 & 0.007&0.007 &0.008 &0.008\\ \cline{2-7}
                        & 1b &0.003 & 0.002&0.002 & 0.002&0.002 \\ \cline{2-7}
                        & 2  & 0.06& 0.07&  0.06& 0.06& 0.07\\ \cline{2-7}
                        & 3 &0.004 &0.005 &0.004 &N/A & N/A\\ \hline
\multirow{4}{*}{\centering \textcolor{black}{Baseline Time (s)}} 
                        & 1a &0.005 &0.12 &0.005 &2.49 & 1.49 \\ \cline{2-7}
                        & 1b &0.003  &0.07&0.002 &2.19 &1.32  \\ \cline{2-7}
                        & 2  & 0.04& 0.92& 0.02& 4.32&1.62 \\ \cline{2-7}
                        & 3  &0.002 &0.02 &0.001 & N/A& N/A\\ \hline
\end{tabular}
\label{tab:computational_cost_analysis}
\end{table}

\credreview{
Execution times vary across classifiers due to their complexity, with deep learning models like WN and DWM taking longer (2.49 and 1.49 seconds on test site 1a) than simpler models like SIC and LR (0.005 seconds on test site 1a).
Notably, and given the nature of Bayesian recursion explained in~\refsection{sec:algorithm}, the computational cost of recursion is independent of the complexity of the underlying instantaneous classifier. For instance, across test site 1a, the recursion times for all models are similar, ranging from 0.007 to 0.008 seconds. 
The cost of recursion increases, however, with image size. For example, test site 2, with the largest images (927×2041 pixels), has recursion times of 0.06 to 0.07 seconds, while test site 1b, with smaller regions (150×110 pixels), shows recursion times as low as 0.002 to 0.003 seconds.
The minimal overhead introduced by recursion is mostly evident for more complex models like RWN and RDWM. For example, the recursion time for RWN constitutes only 0.32\% of the baseline time on test site 1a, 0.09\% on test site 1b, and 1.39\% on test site 2. Similarly, for RDWM, the recursion time is 0.54\% on test site 1a, 0.15\% on test site 1b, and 4.32\% on test site 2.
Moreover, given the nature of Bayesian recursion introduced in~\refsection{sec:algorithm} the RBC cost at each time step remains constant regardless of the time-series length. This reinforces its suitability for real-time remote sensing applications.
}

%
%
\section{Discussion}
\label{section:discussion}
\credreview{
This section discusses the implications of the main findings from this research. For a quantitative summary of the results, please refer to the conclusion in~\refsection{section:conclusion}.}

\subsection{Study Overview and Significance}
\credreview{
Real-time updating of classification maps in remote sensing remains a significant challenge, particularly when training data is limited. This article presents the RBC framework as a solution that enhances robustness and accuracy in land cover classification across multitemporal settings. By using information from previous time steps, RBC successfully handles disturbances present in remote sensing imagery such as illumination and atmospheric interference, e.g., different aerosol concentrations or viewing angles. It can be applied atop any instantaneous classifier based on either a generative or discriminative model.
Furthermore, classification results obtained with a three-class experiment using data from test site 2, as presented in~\refappendix{appendix:charles_river}, illustrate the scalability of the RBC framework to more complex classification tasks.
}

\credreview{
This paper also introduces the SIC classifier, which converts a spectral index value into a probability measure using the mapping in~\refequation{eq:methods:scaled_index_model_conditional_PMF} (please refer to~\reffigure{fig:sic_diagram}). Spectral indices are highly regarded in the remote sensing community owing to their simplicity, interpretability and low computational cost. However, they are often sensitive to changes in illumination and pixel disturbances. Applying RBC on top of SIC helps to address these difficulties and improves the overall strength of the classification.}

\credreview{
The versatility of RBC is shown in its application to various models, ranging from traditional algorithms for machine learning like GMM and LR to sophisticated deep learning models such as DeepWaterMap and WatNet.
Deep learning models, such as deep neural networks, offer great flexibility but can sometimes lead to overconfident classification results. This overconfidence reduces the impact of information from previous time steps and therefore compromises algorithm robustness. To circumvent this phenomenon, we propose to empirically reduce such overconfidence by inserting a positive constant to slightly push the probabilities towards a discrete uniform distribution as in~\refequation{eq:normalization}.
}

\credreview{
Reliability and low computational overhead (as demonstrated in~\refsection{sec:computational_cost_analysis}) make RBC an excellent choice for real-time remote sensing applications where fast updates are essential, e.g., environmental monitoring and disaster management. RBC can enable quick decision-making in the presence of events like deforestation, wildfires, and floods. Being able to manage such diverse and critical tasks highlights the broad importance and potential influence of the proposed framework in many real-world applications.}

\credreview{
RBC is simple, easy to use, interpretable, and controlled by a unique explainable hyperparameter $\epsilon$ which regulates the probability of transitioning among the different classification labels. As a consequence, $\epsilon$ governs the trade-off between adaptability to natural changes in the scene and robustness to outliers caused by illumination of atmospheric interferences.
As the class transition probability is specific to each site and the instantaneous classification algorithm, a study has been conducted on the model sensitivity to the hyperparameter $\epsilon$ in the context of water mapping and deforestation detection. The following subsection looks into the process of selecting this hyperparameter.
}

\subsection{Hyperparameter Selection}
\credreview{
The scarcity of labeled temporal data makes unsupervised strategies particularly appealing for the parameter selection task. One potential approach is to maximize the model evidence or marginal likelihood of the test data~\citep{barberBRML2011}. While this method aligns with Bayesian principles, it requires further investigation beyond the scope of this study.
Therefore, based on the insights gained from the sensitivity analysis in~\refsection{section:sensitivity_analysis}, we propose the following guidelines for selecting $\epsilon$.
}

\credreview{
Overall, the analysis shows that to obtain robust classification results, i.e., that are insensitive to undesired abrupt changes in the image, it is best to select small values of $\epsilon$.
In the absence of ground truth data, we propose to first assess the evolution of the number of pixels classified as a specific class over time for different \(\epsilon\) values. For instance, in a water mapping experiment, tracking how the number of pixels classified as water changes over time under different \(\epsilon\) values may be informative. The sensitivity of the scene to \(\epsilon\) is best evaluated on dates with significant changes in class distribution, such as those caused by natural phenomena like draining events in preparation for extreme rainfall in the water mapping task.
The selection process begins by determining the appropriate increment between tested \(\epsilon\) values to observe changes in the algorithm behavior. Starting with an \(\epsilon\) value close to 0 offers high robustness, and the value can be gradually increased in increments of 0.005. If variability is not observed in the curves showing the number of water pixels over time, the increment can be doubled until changes are detected.
From our experience, effective \(\epsilon\) values typically range between 0 and 0.1, as these values are small enough to ensure robustness while still allowing for adaptability. Importantly, $\epsilon$ values greater than 0.5 should not be tested, as they suggest the typically unrealistic scenario where a pixel is more likely to transition between classes than remain in the same class.
When there is a change in the scene, such as draining, the selected \(\epsilon\) value should allow the model to reflect this change smoothly. This means \(\epsilon\) should be sensitive enough to capture the change, yet robust enough to avoid erratic fluctuations, ensuring both adaptability and stability. If multiple \(\epsilon\) values achieve this balance, the final choice can be made arbitrarily among them.
}

\subsection{RBC framework limitations and challenges}
\credreview{
\label{sec:discussion_limitations}
In this subsection, we discuss the limitations of the proposed framework. These include the lack of ground truth data, which challenges quantitative performance evaluation; missing data due to cloud cover and other atmospheric disturbances; reliance on constant class transition and prior probabilities; and the omission of spatial correlations between classification results across different pixels.}

\credreview{
The main limitation of this work is the scarcity of ground truth multitemporal classification maps, which complicates the quantitative assessment of results. For test sites 1a, 1b, and 2, the authors manually labeled a water mapping dataset with the LabelStudio tool. The labeled dataset has been shared with the remote sensing community to support researchers facing similar challenges~\citep{zenodo_link}. This labeling process is both time-consuming and resource-intensive, which limits the extension of the evaluation of the proposed framework under multiple geographic and environmental conditions.
The availability of open-source labeled deforestation data in the MultiEarth challenge dataset~\citep{cha2023multiearth} facilitated a quantitative analysis for test site 3.
However, this dataset lacks deforestation labels for each acquisition date, allowing quantitative metrics to be computed for only a limited portion of the time series. Specifically, error classification maps were generated for just five images, indicated by filled blue markers in~\reffigure{fig:timeline}. Additionally, temporal misalignment between image and label dates in the MultiEarth dataset required comparing each label to the nearest classification result in time, which introduced temporal gaps.
}

\credreview{
Despite the limited availability of temporal datasets with ground truth, we evaluated our algorithm under a variety of settings to demonstrate its full capabilities, versatility, and potential limitations. We carefully selected three geographical regions: the Oroville Dam (California, USA), Charles River (Boston, USA), and the Amazon rainforest (Brazil), each presenting unique challenges (see~\reffigure{fig:data_aos:challenges}). Cloud cover and shadows were particularly problematic in the Amazon rainforest, where we used CI$_2$ and CSI indices to assess cloud contamination~\citep{zhai_cloud_shadow_detector}. On specific dates, such as 2020-06-10 and 2020-08-04, the presence of clouds caused interference, but RBC was able to improve classification accuracy. Significant differences in illumination were observed between the winter of 2020 and the fall of 2021 in the Oroville Dam region, especially at test site 1a. Additionally, natural events such as substantial fluctuations in water levels at Oroville Dam, particularly at test site 1b from April 2021 to September 2021, presented unforeseen changes that challenged the framework. The summer of 2021 also saw the emergence of seasonal algal blooms in the Charles River, creating some of the most challenging conditions.
}

\credreview{
RBC effectively handles temporary disruptions in remote sensing imagery and resolves missing data from cloud cover by using information from previous time steps. However, prolonged cloud cover and persistent disruptions can negatively impacts classification performance. In general, extended cloud cover may hinder the overall efficiency of the classification process with RBC.
During data preprocessing for test site 3, 182 of the 225 available images were discarded due to high cloud and cloud shadow percentages as indicated by the CI$_2$ and CSI indices, and an additional 12 images were filtered out through visual inspection.
Although this level of cloud cover did not deny the performance of RBC and allowed us to demonstrate its capability to handle temporary disruptions, continuous cloud cover and a larger extent of missing data would need additional strategies.
Temporal interpolation is a potential solution that can be employed to estimate absent values using data from other periods to ensure dataset continuity. However, a recent investigation~\citep{CHE202473} that used linear interpolation to fill in missing values in time series data showed only marginal enhancements in classification accuracy when deep learning models were used. The authors emphasized the need for stronger methods to deal with persistent gaps in data, such as data fusion from multiple sensors and RNNs with masking capabilities. Another recent study uses the Whittaker smoothing to reduce noise and produce continuous NDVI time series using satellite data collected from multiple sources. This technique benefits the precision and spatial resolution of the reconstructed images~\citep{liang2023using}.
Consequently, further research is needed to enhance the adaptability of RBC to diverse environments and develop improved strategies for managing larger levels of missing data.
}

\credreview{
Our methodology, which relies on constant transition probabilities, is computationally efficient, scalable across large geographic areas, and relatively easy to tune. However, this simplicity may come at the cost of accurately reflecting real-world dynamics, as natural systems and human activities often experience shifts that influence transition probabilities. Incorporating recursion could help automatically determine these transition probabilities over time, and a similar approach could be applied to class prior probabilities.
Another limitation of our work is the absence of spatial correlation between classification results at different pixels. By assuming independence between pixel labels, the RBC framework fails to leverage the spatial structures and patterns that are prevalent in natural environments. Addressing these limitations could lead to meaningful advancements in future research.
}

%
%
\section{Conclusion}\label{section:conclusion}
\label{section:conclusion}
\credreview{In this paper, we have introduced the recursive Bayesian classifier (RBC), a framework that converts any instantaneous classifier into a robust online method through a probabilistic approach that is resilient to non-informative image variations.
Using Sentinel-2 data, we have applied RBC to GMM, LR, and to our proposed SIC algorithm, which uses standard broadband spectral indices to generate predictive probabilities. The conducted experiments involve water mapping of the Oroville Dam in California, the Charles River basin in Massachusetts, and deforestation detection in the Amazon.
The results demonstrate that RBC significantly improves the robustness of classifiers in multitemporal settings under challenging conditions. Specifically, in water mapping, RBC enhances balanced classification accuracy by up to 26.95\% for SIC, 13.81\% for LR, and 12.4\% for GMM. In deforestation detection, the accuracy improvements are 15.25\% for SIC, 14.17\% for GMM, and 14.7\% for LR. Additionally, without requiring additional training data, RBC improves the performance of state-of-the-art deep learning models, with DeepWaterMap showing a 9.62\% accuracy increase and WatNet improving by 11.03\% thanks to recursion.
Despite these significant gains, RBC introduces a low computational cost, with minimal overhead that can be considered negligible when applied to more complex deep learning models. For instance, recursion time for recursive WatNet constitutes only 0.32\% of the total algorithm time on test site 1a, 0.09\% on test site 1b, and 1.39\% on test site 2. Similarly, for recursive DeepWaterMap, the recursion overhead is 0.54\% on test site 1a, 0.15\% on test site 1b, and 4.32\% on test site 2. This overhead, in addition to being small, remains constant for each time step regardless of the image time-series length due to the nature of Bayesian recursion, making the proposed methodology a suitable solution for real-time remote sensing applications.
Future work will focus on methods for automatically determining class transition probabilities and addressing the issue of missing data caused by cloud cover and other disruptions in satellite imagery.}

\section*{Declaration of Competing Interest}
The authors of this paper declare that they have no known personal relationships or competing financial interests that could have appeared to influence this research.
\section*{Dataset and supplemental results}
A Python implementation of the proposed algorithms can be found at~\url{https://github.com/neu-spiral/RBC-SatImg}. The pre-processed data \cred{and the manually generated ground truth labels for the water mapping experiments} are available at~\citep{zenodo_link}. Supplemental material containing additional experimental results is also available with this paper.
\section*{Acknowledgements}
This work has been partially supported by the National Geographic Society under grant NGS-86713T-21, the National Science Foundation under Awards ECCS-1845833 and CCF-2326559, and the National Aeronautics and Space Administration under Award 80NSSC20K0742.

\begin{appendices}
\section{Land cover classification experiment}
\label{appendix:charles_river}

\cred{Land cover classification results for test site 2 are presented in~\reffigure{fig:results:Site_c_results}. Performance evaluation in this experiment is restricted to the visual inspection of the classification maps given the lack of ground truth data for the vegetation class in this test site.}
\cred{Results suggest that seasonal variations in the distribution of land, water and vegetation are captured well by the instantaneous classifiers and their recursive counterparts. The decrease in the amount of vegetation starting from November (through winter) with an increase in dry land (at dates 2020-12-13 and 2021-03-20) is represented by an increase in yellow pixels until May, followed by an increase in the number of vegetation pixels through summer and fall (from 2021-05-27 to 2021-09-14).}
\cred{The advantages of the RBC framework are evident, especially on the date 2020-12-13, where the instantaneous classifiers misclassify a substantial land area as water, while the recursive algorithms properly identify the land pixels. On 2021-09-14, both the SIC and LR algorithms fail to identify a section of the water body, yet their recursive counterparts successfully handle this task.}
%

\begin{figure}[h!]
\centering
{\includegraphics[width=14cm]{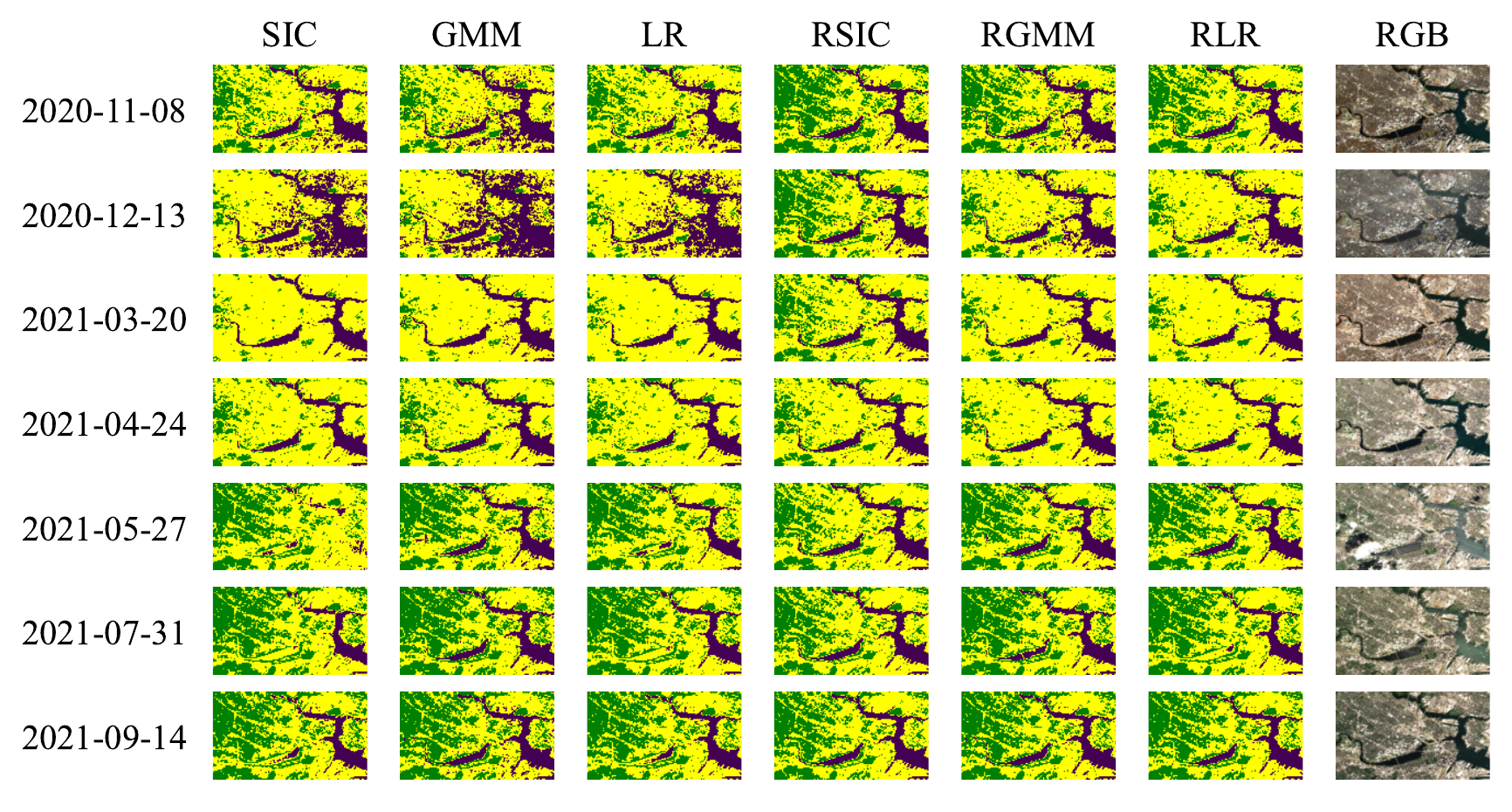}}
\caption{\cred{Land cover classification map results obtained for the Charles river basin area covered by test site 2. Purple, green and yellow represent water, vegetation and land, respectively. Images are arranged in chronological order. By visually inspecting these results, it can be assessed that the proposed RBC framework provides adaptability to seasonal changes and robustness to highly reflective surfaces and other disturbances.}}
\label{fig:results:Site_c_results}
\end{figure}

\arrayrulecolor{black} 
\begin{table}[h]
\small
\centering
    \caption{\cred{Parameter settings for the three-class land cover classification experiment with data from test site 2.}}\label{tab:parameter_settings_appendix}
\begin{tabular}{|l|ll|}
\hline
\multirow{7}{*}{Test site 2} 

& 
\multicolumn{2}{l|}{$C_t \in \mathscr{C}_\mathrm{LC}=\{\mathsf{water},\ \mathsf{land},\ \mathsf{vegetation}\}$}\\ 

\cline{2-3}&\multicolumn{1}{l|}{\multirow{3}{*}{SIC}}& 
$\boldsymbol{\tau}_\mathrm{LC} = [-1,\ -0.05,\ 0.35,\ 1]$;\\
&\multicolumn{1}{l|}{\multirow{2}{*}{}} & $\boldsymbol{\mu}_\mathrm{LC}=[-0.525,\ 0.149,\ 0.675]$;\\
&\multicolumn{1}{l|}{\multirow{2}{*}{}} &$\boldsymbol{\sigma}_\mathrm{LC}=[0.475,\ 0.19,\ 0.325]$;\\
&\multicolumn{1}{l|}{\multirow{2}{*}{}} &$y_{\rm{NDVI}}\left(\bz_t\right) = \frac{z_{t,\rm{NIR}} - z_{t,\rm{red}}}{z_{t,\rm{NIR}}  +z_{t,\rm{red}}}$ \\  

\cline{2-3} 
                               & \multicolumn{1}{l|}{RSIC}                      & $\epsilon=0.05$; $\lambda=0 $    \\ \cline{2-3} 
                               & \multicolumn{1}{l|}{RGMM}                      & $\epsilon=0.05$; $\lambda=0  $         \\ 
                             \cline{2-3} 
                               & \multicolumn{1}{l|}{RLR}                      & $\epsilon=0.05$; $\lambda=0  $   \\
                               \hline
\end{tabular}
\end{table}


\end{appendices}

%
\bibliographystyle{elsarticle-harv}
\bibliography{main}
%
%

\section*{Supplemental Results}
%
%
\begin{figure}[H]
    \hspace{2.4cm}
    \includegraphics[width=17cm]{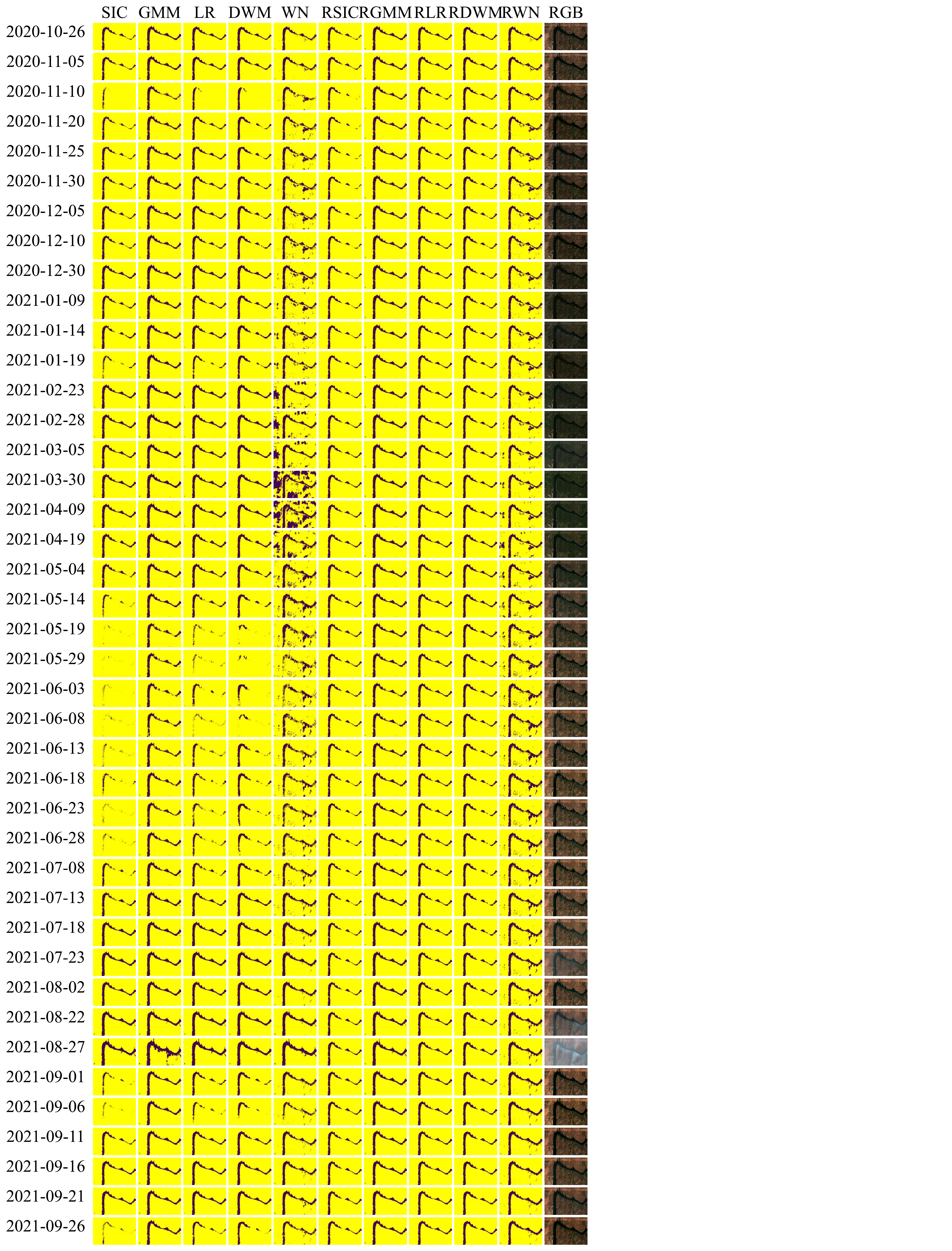}
    \caption{Supplemental results from test site 1a including classification maps for all dates (from 2020-10-26 to 2021-09-26).}
\end{figure}
%
%
%
%
\begin{figure}[H]
    \hspace{2cm}
    \includegraphics[width=18cm]{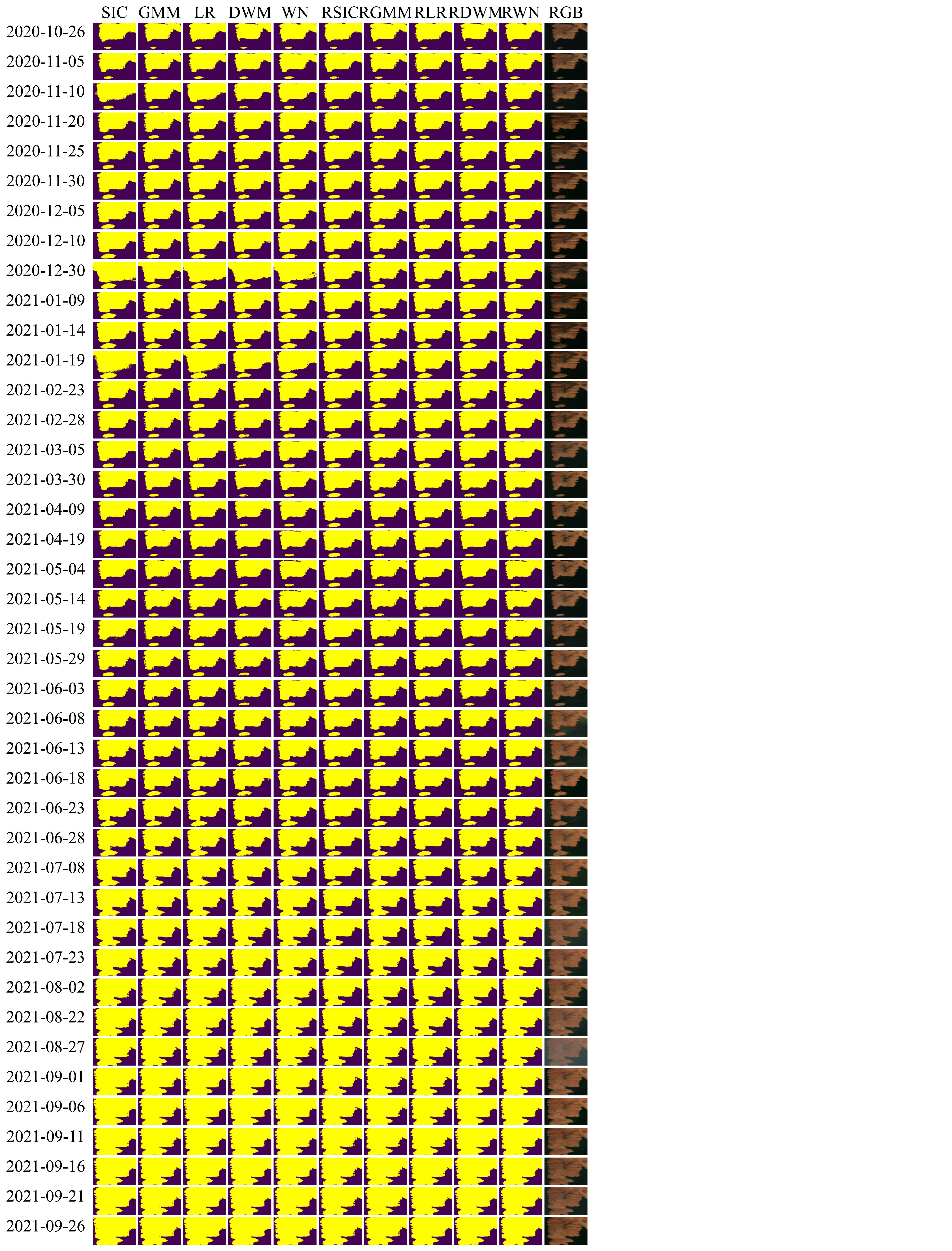}
    \caption{Supplemental results from test site 1b including classification maps for all dates (from 2020-10-26 to 2021-09-26).}
\end{figure}
%
%
%
%
\begin{figure}[ht]
\centering
{\includegraphics[width=16cm]{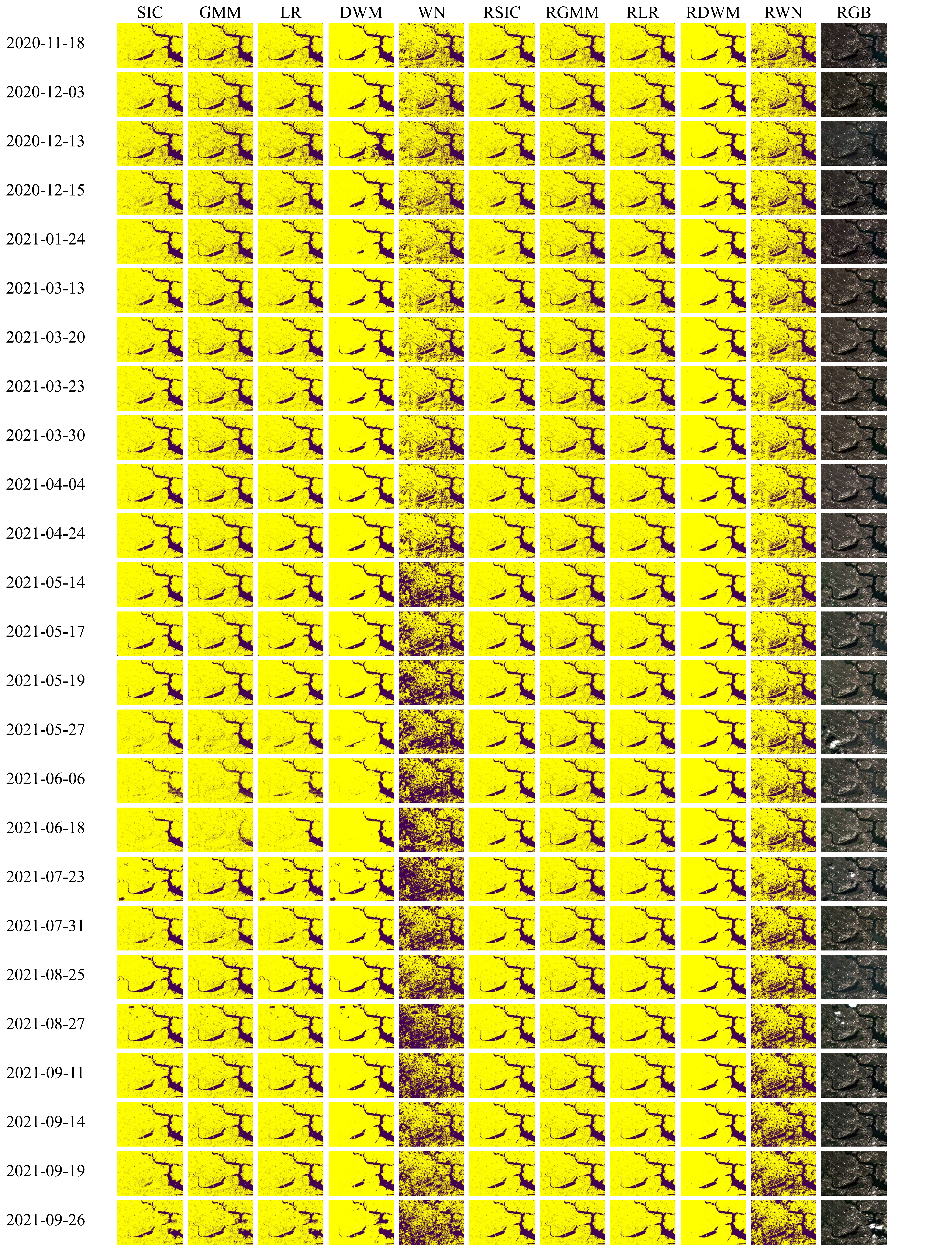}}
\caption{Supplemental results from test site 2 including classification maps for all dates (from 2020-11-18 to 2021-09-26).}
\end{figure}

%
%
\begin{figure}[ht]
\hspace{3cm}
{\includegraphics[width=18cm]{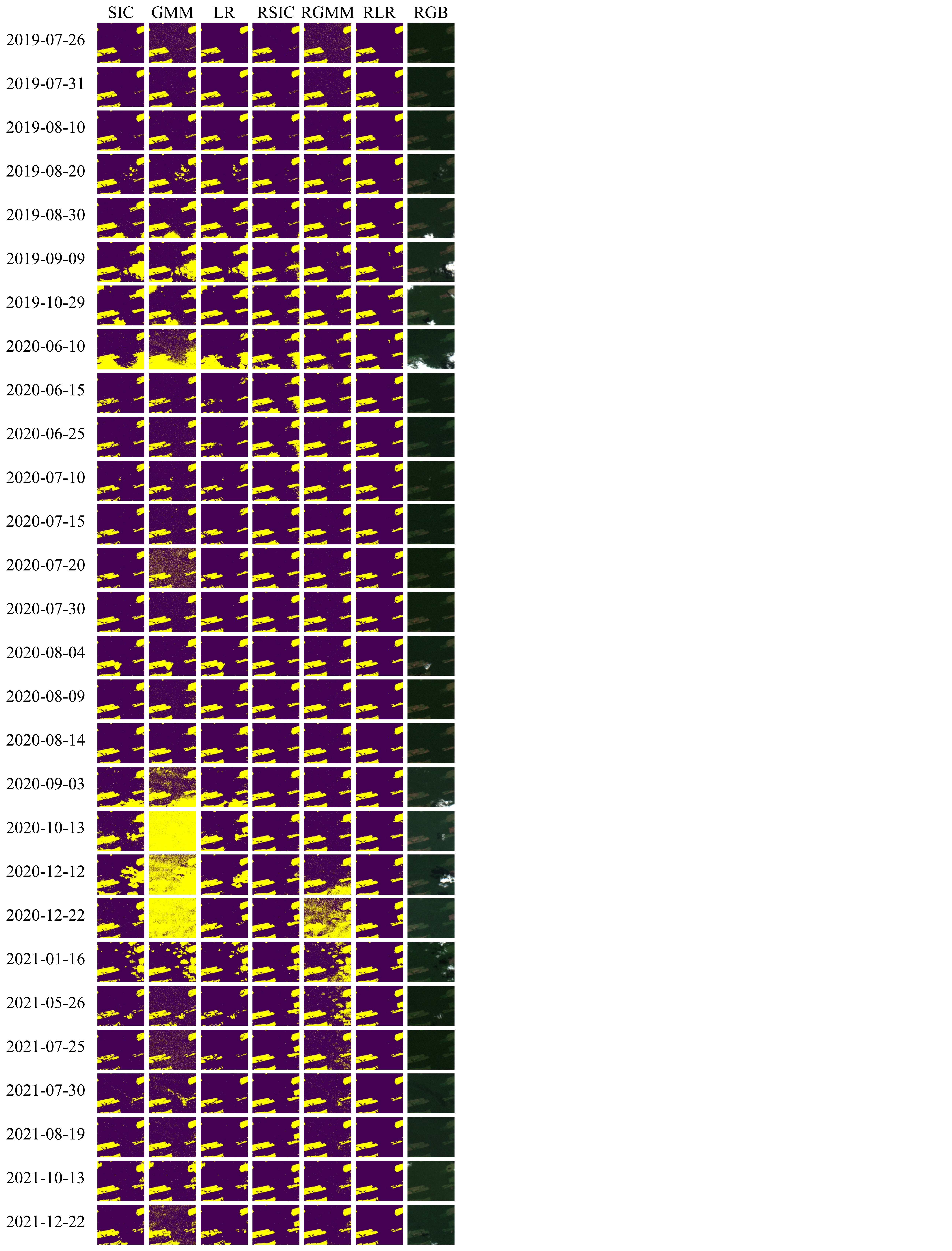}}
\caption{Supplemental results from test site 3 including classification maps for all dates (from 2019-07-26 to 2021-12-22).}
\end{figure}
%
%
%
%
%
\end{document}